\documentclass[aps,prb,twocolumn,%
nolongbibliography]{revtex4-2}
\usepackage[T1]{fontenc}
\usepackage[utf8]{inputenc}
\usepackage{amsmath}
\usepackage{amsfonts}
\usepackage{amssymb}
\usepackage{bm}
\usepackage{graphicx}
\usepackage{hyperref}
\usepackage{mathtools}
\usepackage{physics}
\usepackage{siunitx}
\usepackage[svgnames]{xcolor}
\usepackage{tikz}
\newcommand{\beq}{\begin{equation}}
\newcommand{\eeq}{\end{equation}}
\newcommand{\bse}{\begin{subequations}}
\newcommand{\ese}{\end{subequations}}
\newcommand{\bea}{\begin{eqnarray}}
\newcommand{\eea}{\end{eqnarray}}
\newcommand{\bem}{\begin{displaymath}}
\newcommand{\eem}{\end{displaymath}}
\newcommand{\bmat}{\begin{bmatrix}}
\newcommand{\ebmat}{\end{bmatrix}}

\newcommand{\bs}{\boldsymbol}
\newcommand{\bc}{\begin{center}}
\newcommand{\ec}{\end{center}}
\newcommand{\bmk}{\bm{k}}
\newcommand{\bmr}{\bm{r}}

\begin{document}
	\title{Moir\'{e} fractals in twisted graphene layers}
	
	\author{Deepanshu Aggarwal}
	\affiliation{Department of Physics,
		Indian Institute of Technology Delhi,
		Hauz Khas,New Delhi 110016}
	
	\author{Rohit Narula}
	\affiliation{Department of Physics,
		Indian Institute of Technology Delhi,
		Hauz Khas,New Delhi 110016}
	
	\author{Sankalpa Ghosh}
	\affiliation{Department of Physics,
		Indian Institute of Technology Delhi,
	Hauz Khas,New Delhi 110016}
	
	\date{\today}
	
	 \begin{abstract}
	 Twisted bilayer graphene (TBLG) subject to a sequence of commensurate external periodic potentials reveals the formation of moir\'{e} fractals (MF) that share striking similarities with the central place theory (CPT) of economic geography, thus uncovering a remarkable connection between twistronics and the geometry of economic zones. MFs arise from the self-similarity of the emergent hierarchy of Brillouin zones (BZ), forming a nested subband structure within the bandwidth of the original moir\'{e} bands. We derive the fractal generators (FG) for TBLG under these external potentials and explore their impact on the hierarchy of the BZ edges and the wavefunctions at the Dirac point. By examining realistic super-moir\'{e} structures (SMS) and demonstrating their equivalence to MFs with periodic perturbations under specific conditions, we establish MFs as a general description for such systems. Furthermore, we uncover parallels between the modification of the BZ hierarchy and magnetic BZ formation in Hofstadter's butterfly (HB), allowing us to construct an incommensurability measure for MFs \textit{vs.} twist angle. The resulting bandstructure hierarchy bolsters correlation effects, pushing more bands within the same energy window for both commensurate and incommensurate TBLG.
	 \end{abstract}

	\maketitle

\section{Introduction}
\begin{figure}[b]
	\centering
	\includegraphics[scale=0.42]{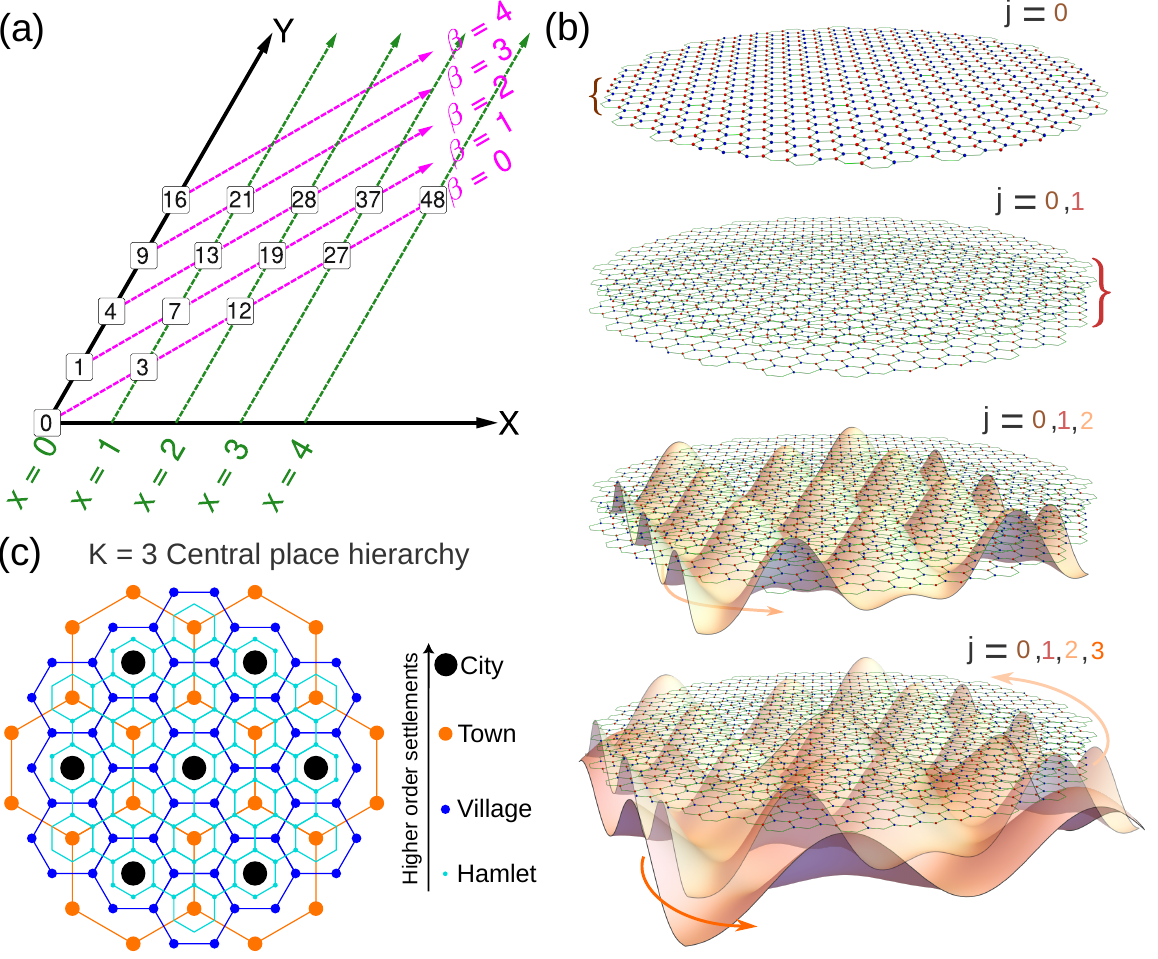}
	\caption{\label{fig:moire_succ_pot} (a) $ L_{N},\beta $ and $x$ over triangular coordinates (the axes are $ \ang{60} $ \textit{w.r.t.} each other).
	The intersection of  $\beta$ (magenta)- and $x$ (green)-lines identify the
intersection points ($x, y=x + \beta$) at which $L_{N}$ exists. (b) The real-space creation of each iteration: TBLG is created by stacking graphene layers with the top layer being the zeroth iteration $j=0$ and the bottom layer being the next $j=1$. The iterations $ j=2,3,\dots $ are created by applying the $z$-independent external periodic potentials identically to both the graphene layers. (c) The $K=3$ CPT hierarchy (since a hexagon of each layer encloses three hexagons of an adjacent layer) where each layer corresponds to a particular order of settlement \cite{Kitchin2009}. $L_{N}$ is analogous to $K$.}
\end{figure}

	Fractals are fascinating structures that are found in both natural and abstract forms, from the intricate patterns of Romanesco broccoli to the complex geometry of the Mandelbrot set \cite{Mandelbrot1983,feder1988}. Iterated function systems (IFS) are a powerful tool for generating fractals, with many unusual geometries emerging as attractors \cite{Duvall1992}, \textit{e.g.,} Koch curve \cite{Lauwerier1991fractals}. Iterated fractals (IF) involve applying a generator recursively to a starting shape --an initiator. Particularly interesting are IFs generated by,
	\begin{equation}
		x^{2} + \beta x - \left(L_{N}-\beta^{2}\right)/3 = 0,
		\label{eqn1}
	\end{equation}
	where for $ \beta \in \mathbb{N}$ if it has a discriminant $ \mathcal{D} \in \mathbb{Z} $, then $ L_{N} \in \mathbb{N} $ generate a triangular lattice with integral coordinates \cite{Dacey1964} (FIG.\ref{fig:moire_succ_pot}(a)).

\begin{figure*}[t]
	\centering
	\includegraphics[scale=0.35]{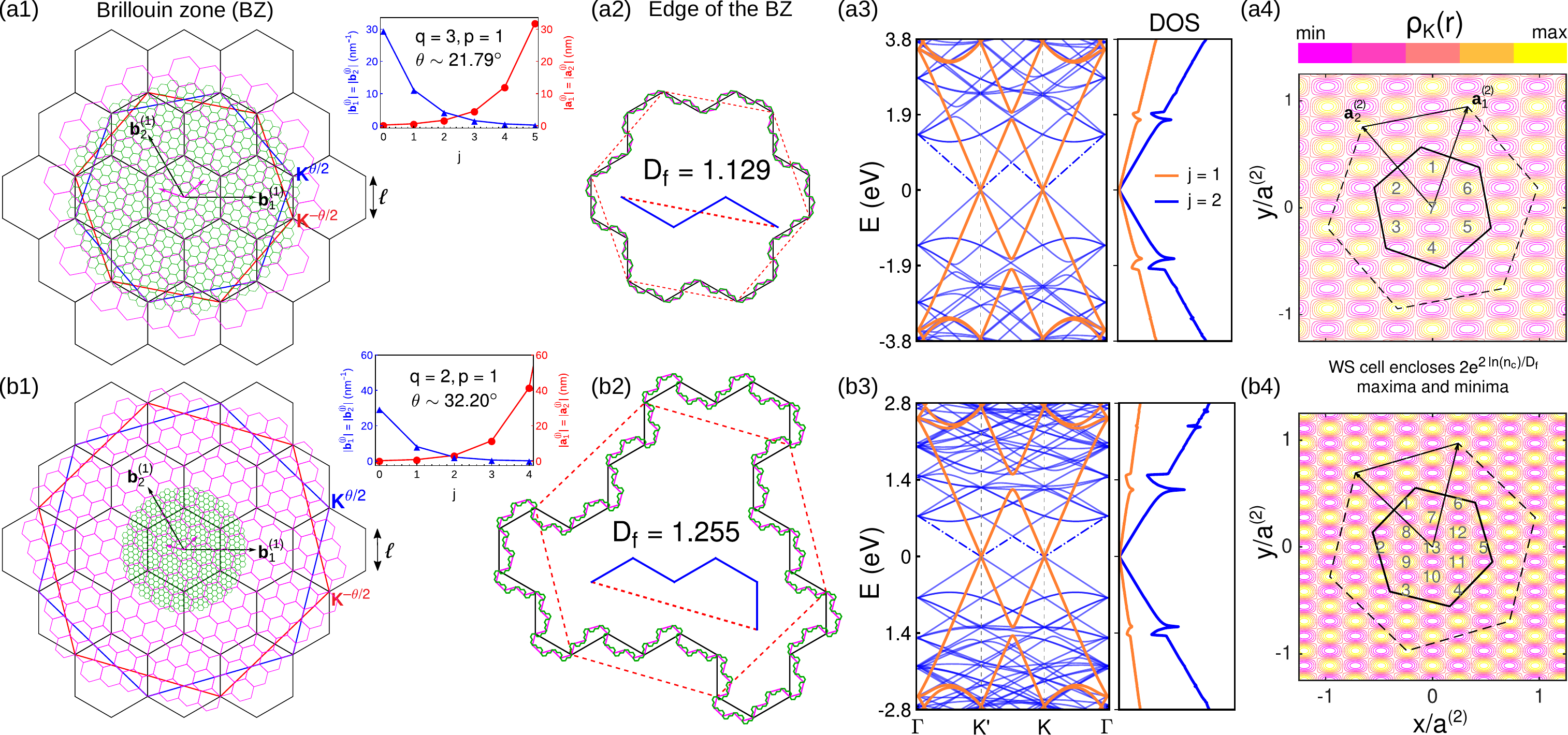}
	\caption{(a1), (b1) The BZ for $ j=1$ (\text{black}), $2$ (\text{magenta}) $3$ (\text{green}), where $\left\{\bm{b}^{(1)}_{1},\bm{b}^{(1)}_{2}\right\}$ are the reciprocal lattice primitive vectors for $j=1$. (a2), (b2) The fractal structures at the BZ edges. The dashed red hexagon represents the initiator, \textit{i.e.}, the BZ of SLG rotated by $\theta/2$. The solid blue lines outline the fractal structures' outer boundary. The generators (inside) attach alternately to the initiator, forming the fractal structure. The copies of the BZ at each iteration are added such that they overlap with the BZ of SLG (red solid line in (a1, b1)) and these overlaps lead to IFs. The insets between (a1), (a2) and (b1), (b2) display reciprocal and real-space lattice vectors for $q=3, p=1$ and $q=2, p=1$, respectively. The shift between the Dirac points is equal to the hexagon's side length for $q=3, p=1$, and twice its side for $q=2, p=1$. (a3) and (b3) The bandstructures for $ V_{0} = 1.2\,\unit{\milli\electronvolt} $ and the DOS (with a Gaussian smearing of $ 0.002\,\unit{\milli\electronvolt} $). (a4) and (b4) show $\rho_{n\bs{k}}(\bmr)$ of the lowest conduction band (dashed-dotted line) at the Dirac point (see text, Appendix-\ref{ssec:8}).}
	\label{fig:fractal_structures_comm}
\end{figure*}

In this work we show that Eq.(\ref{eqn1}) describes an emergent fractality when commensurate or incommensurate moir\'{e} patterns in TBLG \cite{Santos2007,Suarez2010,Li2010,Hicks2011,Bistritzer2011,Santos2012,Lu2013,PMoon2013,Sboy2015,Dai2016,CaoPRL2016,Peeters2018,Huder2018,Qiao2018,Zou2018,Cao2018I,Cao2018II,Carr2018,Chandan2018,Lian2019,Tomarken2019,Carr2019,Wolf2019,Kwang2019,Andrei2020,Balents2020,Zondiner2020,Bern12021,Song2021,TingxinLi2021,LiguoMa2021,Debnath2021,Gardezi2021,Wolf2021,Hesp2021,DAggarwal2023,DArora2023} are subjected to a sequence of superlattice periodic potentials (SOPP) (FIG.\ref{fig:moire_succ_pot}(b)), having the same moir\'{e} periodicity as structures on which they are applied, but twisted by an angle restoring commensuration. The sequence of iterated edges of the first Brillouin zone (FBZ) forms a fractal (FIG.\ref{fig:fractal_structures_comm}, \ref{fig:frac_strs_incomm}) with dimensions determined by $L_{N}$ (the number of unit cells in a newly formed BZ fitting a unit cell of the preceding BZ at each iteration).
	
The emergent fractality of Eq.(\ref{eqn1}), dubbed as the moir\'e fractal (MF) resembles the hierarchy and fractality of economic geography's CPT as pioneered by Christaller \cite{Christaller1966central,Arthur1966,WC1983} and L\"{o}sch \cite{loschbook}, which terms $L_{N}$ as L\"oschian numbers \cite{Marshall1975}. This connection emerges when densely-packed hexagonal trade areas centered on settlements are multiply-stacked with trade areas representing smaller settlements \cite{WC1983,loschbook,Cardillo2006,Courtat2011,Barthelemy2008,Barthelemy2013,Barthelemy2019,HongZhang2022} (FIG.\ref{fig:moire_succ_pot}(c)). The MF fractal dimension ($D_f$) provides quantitative information about the bandstructure of realistic SMS \textit{e.g.,} multiple graphene or hexagonal boron nitride (hBN)-graphene layers \cite{Bernevig2019,Kaxiras2020,Aviuri2023superconductivity,Liu2019,Xie2022,Francois2020}. Further, we establish an analogy with HB \cite{Hofs1976,Albrecht2001} explaining the topological quantization of Hall conductivity \cite{Thouless1982,Kohomoto1985}, thus formulating an incommensuration measure for moir\'{e} structures.

\section{The Hamiltonian and emergent fractality in Moir\'{e} Fractals}

The rapid progress in the fabrication of two-dimensional (2D) layered materials \textit{e.g.,} TBLG has stimulated interest in the effect of substrates \cite{Novoselov2004,Novoselov2005pnas,Geim2013,Dean2012}. Such materials experience an external potential with a moir\'{e}-like periodicity when placed on substrates with matching \cite{MKinder2011} or mismatched layers \cite{MKinder2012,Wallbank2013} which may be modeled as perturbations to the Hamiltonian of TBLG ($H_{\text{TBLG}})$ via external periodic potentials \cite{Yankowitz2012,Wallbank2013,Pmoon2014}. $H_{\text{TBLG}}$ subjected to SOPP at $ j $-th iteration is
\begin{equation}
	H_{j} =
	\begin{bmatrix}
		\hat{h}_{\bmk}(\theta/2)+ \sum_{i=1}^{j}V_{i}(\bmr) & T(\bmr) \\ T^{\dagger}(\bmr) & \hat{h}_{\bmk}(-\theta/2) + \sum_{i=1}^{j}V_{i}(\bmr)
	\end{bmatrix}
	\label{eqn:2ham}
\end{equation}
where $\hat{h}_{\bmk}(\theta) = v_{F}\,\bm{\sigma}_{\theta}\vdot\left(\hat{\bmk} - \bm{K}^{\theta}\right)$ \cite{CastroNeto2009} describes single-layer graphene (SLG) rotated by $\theta$, $v_F$ \cite{Sreich2002} is the Fermi velocity, and $\bm{K}^{\theta}$ is the rotated, right-valley Dirac point. The transformed Pauli matrices $\bm{\sigma}_{\theta} = e^{-i\sigma_z \theta/2}(\sigma_x,\sigma_y)e^{i\sigma_z\theta/2}$ provide the rotation. The expressions for the interlayer hopping matrices $T(\bmr)$ \cite{Bistritzer2010,Bistritzer2011,Vishwanath2019,Ledwith2021} are given in appendix-\ref{ssec:6}. The external potential $V_{j}(\bmr)$ exhibits a periodicity: $V_{j}(\bmr + n_1\,\bm{t}^{(j)}_1 + n_2\,\bm{t}^{(j)}_2) = V_{j}(\bmr)$, where the primitive vectors (PV) $\bm{t}^{(j)}_{1,2} = [\mathcal{R}(\theta)]\,\bm{a}^{(j-1)}_{1,2}$. $\theta= \theta_{r}$ is the twist between the moir\'{e} pattern and the moir\'{e} external potential (mEP), leading to commensuration between mEP at each $j$-th iteration and TBLG with all potentials up to the $(j-1)$-th iteration. $\mathcal{R}(\theta_r)$ denotes a 2D rotation matrix at these commensurate angles. The condition for commensuration under such rotation maps an integer pair $\bm{n} = \{n_1,n_2\}$ to $\bm{m} = \{m_1,m_2\}$ (Appendix-\ref{ssec:4}). 
The integral solutions: $m_1,m_2,n_1$ and $n_2$ satisfy the necessary and sufficient condition when the matrix elements assume only rational values \cite{Shallcross2010,Santos2012}, leading to a set of Diophantine equations (Appendix-\ref{ssec:4}) whose solutions provide the PV of the commensurate supercell, \textit{i.e.}, $ \left\{\bm{a}^{(2)}_{1},\bm{a}^{(2)}_{2}\right\} $ in terms of the PV of the preceding structure.

For the SOPP of FIG.\ref{fig:moire_succ_pot}, $ V_{1}(\bm{r}) = 0 $ at  $ j=1 $, $ H_{1} $ in (\ref{eqn:2ham}) becomes $H_{\text{TBLG}}$, while the potentials for $ j\ge 2 $ are non-zero. For $ j=2 $, $ V_{2}(\bmr) $ is periodic with $ \bm{t}^{(2)}_{1,2} = \left[\mathcal{R}(\theta_{r})\right]\bm{a}^{(1))}_{1,2}$ where $ \bm{a}^{(1)}_{1,2} $ are the PV of commensurate TBLG. On repeating this procedure, the commensuration of $V_{j}(\bmr)$ and the structure upto the $ (j-1) $th-level is spanned by $ \left\{\bm{a}^{(j)}_{1},\bm{a}^{(j)}_{2}\right\} $. We demonstrate this via a cosine potential $ V_0 $ with only six Fourier components (Appendix-\ref{ssec:6}).
The intrinsic Coulomb interactions can be modeled using such an onsite mEP having the same periodicity as the moir\'{e} pattern\cite{Guinea2018,Rademaker2019,Goodwin2020,Cao2021moireSupflat} as a starting \textit{ansatz} for a self-consistent calculation. For the first iteration of $V_{2}(\bmr)$ (Appendix-\ref{ssec:7}-FIGs.\ref{fig:layersfig}-\ref{fig:trilayhbnghbn}) various SMS \textit{e.g.,} trilayer graphene \cite{Bernevig2019,Kaxiras2020,Chen2021,He2021,Xu2021,Aviuri2023superconductivity}, four-layer graphene \cite{Liu2019,Burg2022,Zhang2022,Xie2022}, and trilayer hBN-G-hBN \cite{Finney2019,Francois2020} are modeled by (\ref{eqn:2ham}) representing an MF and a weak periodic perturbation whose details depend on the system considered. $V_{0}$ can be controlled by \cite{Garcia2021} the interlayer separation ($d$) and the interlayer bias ($V_{\text{STM}}$), essentially the bias applied to a scanning tunneling microscope (STM) tip, \textit{i.e.,} $\abs{V_{\text{STM}}} \approx 20-500\,\unit{\milli\electronvolt}$ \cite{Choi2019,Nuckolls2023}. Typically, $V_{0} \sim 1.2\,\unit{\milli\electronvolt}$ for $\abs{V_{\text{STM}}} = 45\,\unit{\milli\electronvolt}$ given $d\sim 1\,\unit{\nano\metre}$.

Each commensuration of either TBLG or TBLG plus the mEPs gives (Appendix-\ref{ssec:4})
\begin{equation}
	A^{(j-1)}_{\text{FBZ}}/A^{(j)}_{\text{FBZ}} = L_{N} = p_1^2 + p_2^2 + p_1\,p_2
	\label{eqn:numberN}
\end{equation}
where $ A^{(j)}_{\text{FBZ}} = \abs{\bm{b}^{(j)}_{1}\times\bm{b}^{(j)}_{2}} $ represents the area of the first BZ at $ j $-th iteration and $ \bm{a}^{(j)}_i\cdot\bm{b}^{(j)}_k = 2\pi\delta_{ik}~\forall~i,k =1,2$ and $p_{1}(p,q), p_{2}(p,q) \in \mathbb{Z}^{+}$ (Appendix-\ref{ssec:4}-FIG.\ref{fig:p1p2space}), with $ p,q $ being co-prime numbers. For a hexagonal lattice, $ p_{2}=p_{1} + \beta $ such that $ L_{N} $ with $ \beta = 0 $ lie on the line $X=Y$ while the numbers with $ \beta > 0 $ lie on lines parallel to $X=Y$ (FIG.\ref{fig:moire_succ_pot}(a)), converting Eq.(\ref{eqn:numberN}) into Eq.(\ref{eqn1}), therefore making each $L_{N}$ lie at an intersection of the $x$- and $\beta$-rays (FIG.\ref{fig:moire_succ_pot}(a)).

For each $ L_{N} $ and $ \beta $, the corresponding FG are generated by applying IFS to one side of the initiator $A_{0}$ \cite{Duvall1992}, \textit{i.e.,} the rotated BZ of SLG (a constituting layer in TBLG) (details in appendix-\ref{ssec:1}).
Applying this FG to each arm of $A_{0}$ generates the edges of successive BZs, continuing the recursive process to produce a sequence of BZs.

The hierarchical construct in CPT \textit{i.e.,} $L_{N}$ exhibits successively smaller regions within a trade area at each stage. Number theory \cite{Arlinghaus1985} identifies conditions for $L_{N}$ fulfilling Eq.(\ref{eqn1}). This connection between the CPT lattice partition and Eq.(\ref{eqn1}) facilitates the systematic determination of lattice coordinates for economic zones and the corresponding FG responsible for CPT associated with $L_{N}$. For TBLG in the presence of specific mEPs, we begin with the transformation mappings of the FGs for $ q=3 $,$ p=1 $ and $q=2,p=1$ corresponding to $\theta \sim \ang{21.79}$\cite{Ilani2023} and $ \theta \sim \ang{32.20} $. $D_{f} = \log(n_{c})/\log(s)$ for the attractor $ A $ \cite{Mandelbrot1983} where $s = \sqrt{L_{N}} $ is the contractivity factor (Appendix-\ref{ssec:1}-FIG.\ref{fig:FG3121}).

\section{Implications of the above construction for the band structure} 
FIG.\ref{fig:fractal_structures_comm}(a1),(b1) display the superimposed BZs for $ j=1,2,3 $ corresponding to the hierarchical mEP applied to two commensurate structures at $ \theta =\theta_{r} \sim \ang{21.79} $ and $ \theta =\theta_{r} \sim \ang{32.20} $. The FG shape is shown in the next column. For both, it is applied alternately outside and inside the edges of the initiators (red dashed lines, FIGs.~\ref{fig:fractal_structures_comm}(a2),(b2)), exhibiting emergent fractality with $ D_{f} = 1.129 $ and $ D_{f} = 1.255 $. The real-space fractals corresponding to $ q=3 $,$ p=1 $ are shown in appendix-\ref{ssec:2}-FIG.\ref{fig:realfracqp31}. The RLV magnitudes for $ i=1,2 $ form a Cauchy sequence \cite{barnsley2012} as  $\lim_{j\rightarrow\infty} \left\{\frac{\abs{\bm{b}^{\pm\theta/2}_{i}}}{s^{\left(j-1\right)}}\right\}\rightarrow 0$ whose convergence rates depend on $q,p$ (Solid blue and red lines in the insets of FIG.\ref{fig:fractal_structures_comm}).

\begin{figure}[t]
	\centering
	\includegraphics[scale=0.5]{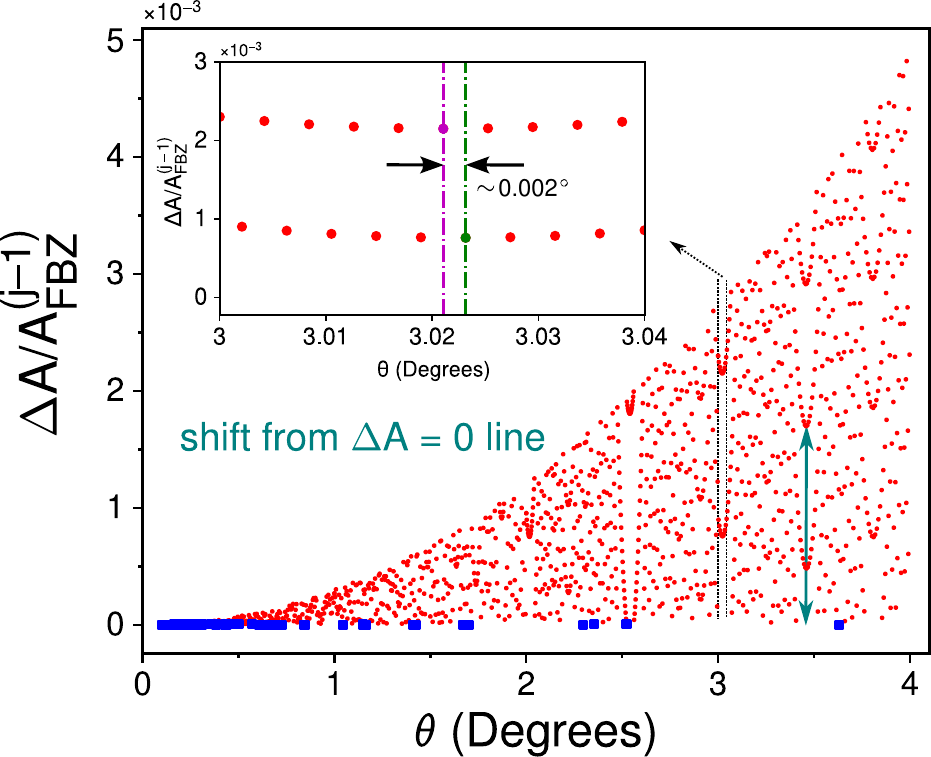}
	\caption{$\Delta A/A^{(j-1)}_{\text{FBZ}}$ \textit{vs.} $\theta$: the blue rectangles represent commensurate/closest angles for odd $q$ and $p=1$. The first BZ area for incommensurate angles is determined using moir\'{e} vectors\cite{Koshino2018}. The inset confirms the absence of the apparent multivaluedness in the lowest values of the V-regions between the dashed lines, with a separation of $\sim \ang{0.002}$.}
        \label{delAplot}
\end{figure}

The commuting translation operators (TOs) at the $j$-th iteration with the Hamiltonian (\ref{eqn:2ham}) for the rotated mEP with PV $ \bm{a}^{(j)}_{1} = p_{1}\bm{a}^{(j-1)}_{1} + p_{2}\bm{a}^{(j-1)}_{2} $ and $ \bm{a}^{(j)}_{2} = -p_{2}\bm{a}^{(j-1)}_{1} + \left(p_{1}+p_{2}\right)\bm{a}^{(j-1)}_{2} $ differ from the PV at the $(j-1)$-th iteration. They satisfy $ \hat{T}_{\bm{a}^{(j)}_{1}}\hat{T}_{\bm{a}^{(j)}_{2}} = \hat{T}_{\bm{a}^{(j)}_{2}} \hat{T}_{\bm{a}^{(j)}_{1}} $, leading to a large real-space supercell with a more squeezed BZ scaled by $s$ (FIG.\ref{fig:fractal_structures_comm}(a1),(b1)), conventionally termed as a  \emph{mini} zone (MZ) \cite{Park2008,Dean2012}. This has similarities with the HB problem \cite{Hofs1976,MacD1983,MacD1984} where the magnetic TOs do not commute. Consequently for a TR-symmetric, commensurate TBLG in a rotated mEP we define a dimensionless incommensurability measure
\footnote{In HB problem, $ \hat{T}_{\bm{a}_{1}}\hat{T}_{\bm{a}_{2}} \ne \hat{T}_{\bm{a}_{2}} \hat{T}_{\bm{a}_{1}} $, where $ \bm{a}_{1} $ and $ \bm{a}_{2} $ are the direct-lattice PV. To ensure commutativity, magnetic TOs are redefined for the enlarged magnetic unit cell. Dimensionless quantity $\Phi/\Phi_{0} = \abs{\bm{a}_{1} \times \bm{a}_{2}}/l^{2}_{B}$ serves as an incommensurability measure with magnetic flux per plaquette $\Phi$,flux quanta $\Phi_{0} = \hbar/e$,and magnetic length $l_{B}$ in the presence of a magnetic field breaking time reversal (TR) symmetry.}
,
\begin{equation}
	\frac{\Delta A (\theta)}{ A^{(j-1)}_{\text{FBZ}}} =  \left( 1 - \frac{\lfloor A^{(j-1)}_{\text{FBZ}}/A^{(j)}_{\text{FBZ}} \rfloor}{A^{(j-1)}_{\text{FBZ}}/A^{(j)}_{\text{FBZ}}} \right) \label{eq:incommen}
\end{equation}
where $\lfloor \dots \rfloor$ denotes the greatest integer function.

Following \cite{Bistritzer2011,Koshino2015,Koshino2018} for any $\theta$ between TBLG and mEP (FIG.\ref{fig:fractal_structures_comm}(b1)), $\bm{b}^{(j)}_{1,2}$ of the mBZ are obtained from $\Delta \bm{K} = \bm{K}^{\theta/2} - \bm{K}^{-\theta/2}$. These definitions coincide for $p=1$ and odd integer $q$ with the lattice vectors
giving the hexagonal BZ side length as $\ell = \abs{\frac{2\bm{b}_{1}^{(1)} + \bm{b}_{2}^{(1)}}{3}}$ (FIG.\ref{fig:fractal_structures_comm}(a1),(b1)), such that $A^{(j-1)}_{\text{FBZ}}/A^{(j)}_{\text{FBZ}}=L_{N} \Rightarrow \Delta A =0$, corresponding to the blue points in FIG.\ref{delAplot}.

For generic $(q,p)$ the two definitions do not coincide and $\Delta A/A^{(j-1)}_{\text{FBZ}}$ shifts upward from the $\Delta A=0$ line by differing amounts (FIG.\ref{delAplot}). The angles corresponding to the minima of the V-regions are commensurate angles where the ratio becomes rational (Appendix-\ref{ssec:5}). FIG.\ref{delAplot}(inset) shows that these minima are not vertically collinear: the lateral shift is $\sim \ang{0.002}$.

FIG.\ref{fig:fractal_structures_comm}(a3),(b3) display the bandstructures for $j=1$ ($H_{\text{TBLG}}$) and $j=2$ in the presence of mEP $V_{2}(\bmr)$, and the DOS. Precisely $2\left(2e^{2\ln(n_{c})/D_{f}}-1\right)$ (Appendix-\ref{ssec:3}) bands, per Eq.\ref{eqn:numberN}, populate the bandgap of $6.38~\si{\electronvolt}$ for $\theta_{r} \sim \ang{21.79}$ and $5.04~\si{\electronvolt}$ for $\theta_{r} \sim \ang{32.20}$ between the two lowest bands at $\Gamma$-point, without the mEP. Additional bands at $K$ and $K'$ occupy an even narrower range within the bandgap \textit{vs.} the $\Gamma$-point, with the two lowest bands meeting at the Dirac point. FIG.\ref{fig:fractal_structures_comm}(a4),(b4) provide the probability density $\rho_{n\bs{k}}=\abs{\psi_{n\bs{k}}}^{2}$ of the Bloch wavefunctions obtained by diagonalizing (\ref{eqn:2ham}) at the Dirac point. The number of maxima and minima within the Wigner-Seitz cell for the lowest conduction band is $2e^{2\ln(n_{c})/D_{f}}$ (Appendix-\ref{ssec:8}). The MFs in (\ref{eqn:2ham}) therefore provide precise control over the number of in-gap states given $D_{f}$ by changing $\theta_{r}$, which is in contrast to other such methods for small-angle TBLG \cite{Jose2013,Efimkin2018,Xu2019} and SMS \cite{ZWang2019,Zhu2020}.

\begin{figure}[b]
	\centering
	\includegraphics[scale=0.36]{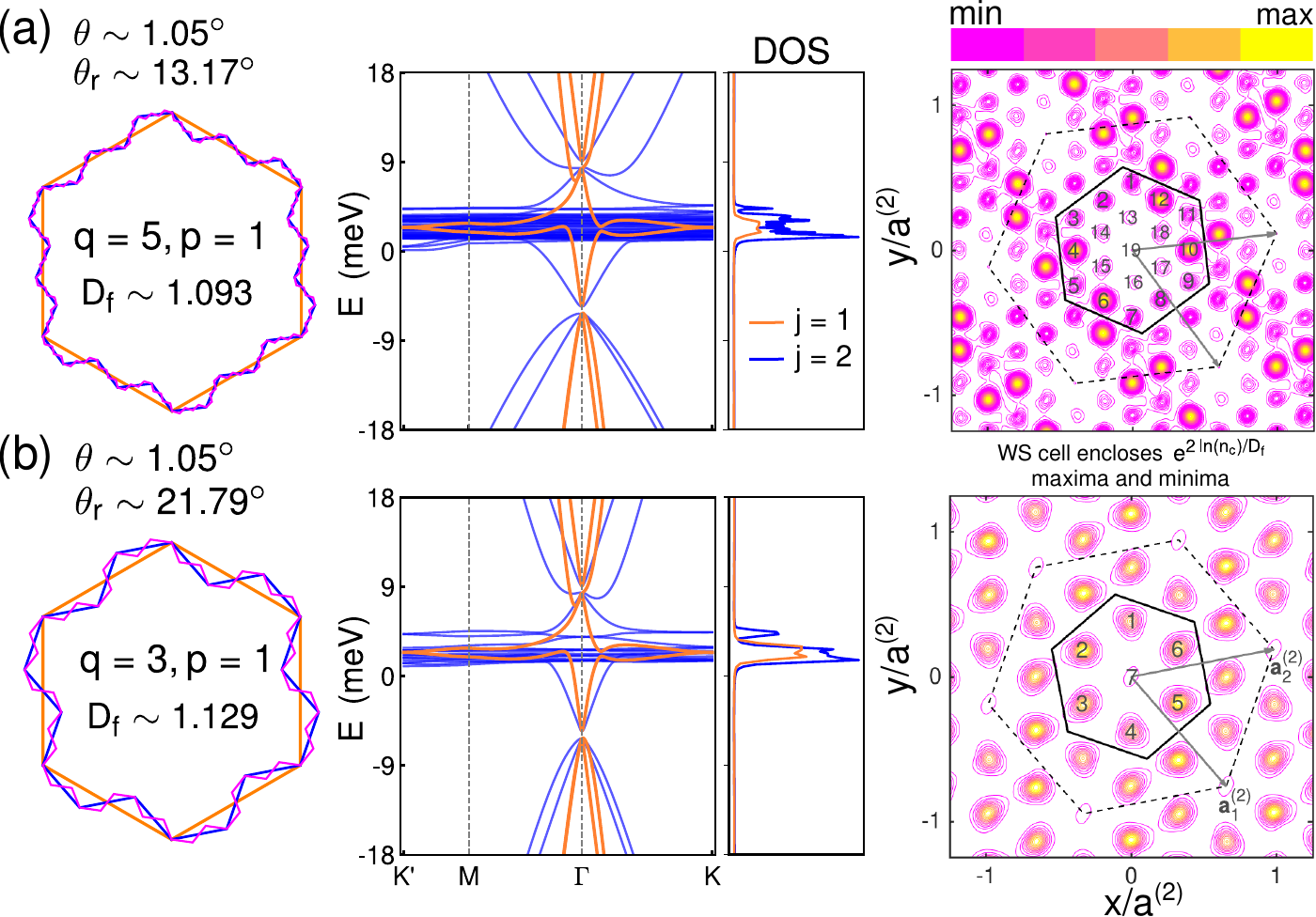}
	\caption{The bandstructures at $ \theta_r \sim \ang{21.79} $ in (a) and $ \theta_r \sim \ang{13.17} $ in (b) along $K'-M-\Gamma-K$. The twist between the layers is the first magic angle $ \theta\sim\ang{1.05}$. The interlayer hopping parameters: $t_{\text{AA/BB}} = t_{AB/BA} = 110~\unit{\milli\electronvolt} $ (Appendix-\ref{ssec:6}) and $V_{0} = 1.2~\unit{\milli\electronvolt}$. (left to right) The emergent fractal structure in reciprocal space, 2D-bandstructure: orange (blue) bands without (with) mEP, the DOS, and the spatial profiles of $\rho_{K}(\bmr)$ of the lowest conduction band with $j=2$. Changing $\theta_{r}$ from (b) to (a) modifies $D_{f}$, shifting more bands towards $E_F$.}
	\label{fig:frac_strs_incomm}
\end{figure}

For magic-angle TBLG (MATBLG) \cite{Suarez2010,Bistritzer2011,Uchida2014,NNTnam2017}, the flat bands facilitate various correlated phases \cite{Cao2018I,Cao2018II,Lu2019,Serlin2020,Nuckolls2020,Wu2021,Saito2021,Das2021,LedwithFCI2020,Cao2021}. The FGs apply alternately outside and inside the edges of their initiators, which are the mBZ of MATBLG (red dashed lines in FIGs.~\ref{fig:frac_strs_incomm}(a),(b)), also showing the superimposed BZs for $j=1,2,3$ at the first magic angle $\theta \sim \ang{1.05}$). For FIG.\ref{fig:frac_strs_incomm}(a): $\theta_{r}\sim\ang{13.17}$ and $D_{f} =1.093$, while for FIG.\ref{fig:frac_strs_incomm}(b): $\theta_{r}\sim\ang{21.79}$ and $D_{f} =1.129$. The change in $D_{f}$ and $L_{N}$ alters the number of bands pushed towards the Fermi level $E_{F}$ at the $\Gamma$-point within a bandgap of $\sim13.76~\si{\milli\electronvolt}$ for MATBLG, preserving the emergent fractality similar to commensurate structures. An additional $2(e^{2\ln(n_{c})/D_{f}}-1)$ inner bands again exhibit significantly reduced curvature compared to the original flat bands. FIG.\ref{fig:Fermivel1} verifies that TBLG's renormalized $v_F$ remains indifferent to the presence of hierarchical mEP, despite the shift in $E_{F}$. The Hamiltonian (\ref{eqn:2ham}) ignores lattice relaxation effects, namely the variations in the interlayer hopping amplitudes in the AA-BB- and AB-BA-rich regions. Their \cite{Koshino2015,Carr2018,Carr2019,Leconte2022} inclusion doesn't alter this emergent fractality linked to the bandstructures (Appendix-\ref{ssec:6}-FIG.\ref{fig:corrband}).

Much like how the recursive stacking of nets in CPT leads to a proliferation of smaller settlements within a designated area, the recursive mEP in TBLG produce an expanding count of electronic bands within a defined energy range, thereby amplifying correlation effects, with both scenarios characterized by $L_{N}$. The impact of this band engineering and emergent fractality on electronic correlation can be understood through the Hubbard parameter ratio between the interaction energy $U$ and the band-width $t_W$ (the difference between the band maxima and minima). Since in (\ref{eqn:2ham}), $H_{\text{TBLG}}$ and the rotated MEP have different translational symmetry, $\lim_{ V_{0} \to 0}\frac{U}{t_{W}}$ doesn't yield the same ratio for pristine TBLG \cite{Cao2018II}. However this ratio depends on mEP $V_{0}$ when $E_{F}$ lies within the flat band, $\frac{U}{t_{W}} \gg 1$. Under an mEP, $U \rightarrow \frac{U}{s}$,for $U = e^{2}\theta/4\pi \kappa\epsilon_{0} a$ without any mEP (Appendix-\ref{ssec:9}), where $e$ is electronic charge and $\kappa=4$. However,the closest flat band near $E_{F}$ experiences a significantly larger reduction in bandwidth \textit{vs.} $s$. \textit{E.g.}, in FIG.\ref{fig:frac_strs_incomm}(a), the bandwidth decreases from $\sim 6\si{\milli\electronvolt}$ to $\sim 0.23\si{\milli\electronvolt}$ (FIG.\ref{fig:bw}), leading to $\sim 9$ times increase in $\frac{U}{t_W}$ \textit{vs.} MATBLG. The effective mass $m^{*}$ scales as $\sqrt{n}/v_{F}$ at $E_{F}$, increasing with the superlattice density $n$ scaled by $L_{N}$; so does DOS.

\section{Conclusion}
In summary, we have introduced MFs in TBLG subjected to a sequence of commensurate, rotated external potentials, thus establishing a framework for studying their properties and drawing parallels between emergent fractality in $sp^{2}$ carbons and the agglomeration of CPT trade zones. The bandstructure of several SMS can be understood via MFs and a weak perturbation enabling insertion of a controlled number of in-gap bands determined by $D_f$. We analyzed the restructuring of the moir\'{e} unit cell and established an incommensurability measure, linking it to correlation effects and $D_f$. The MFs remained robust despite corrugation effects.

Amidst the emerging domain of SMS, including trilayer \cite{Chen2021,He2021,Xu2021,Aviuri2023superconductivity}, tetralayer \cite{Burg2019,Shen2020,He2021double,Burg2022}, and pentalayer graphene \cite{Park2022,Zhang2022}, scenarios where slight rotations of SLG interact with thin graphitic crystals \cite{Mullan2023,Waters2023}, dissimilar layers like encapsulated SLG and bilayer graphene between hBN layers \cite{Finney2019,Lujun2019,Francois2020,Smeyers2023}, and moir\'{e} lattices in photonic crystals \cite{Wang2020} and ultra-cold atom systems \cite{Salamon2020,Luo2021,Meng2023}, we've provided a general framework allowing such systems to be understood as MFs under weak perturbations.

The increase in the number of MF bands near $E_{F}$ can be detected via
differential tunneling conductance measurements for the corresponding SMS under suitable conditions (Appendix-\ref{ssec:10}) as was done for quasicrystals with
a Penrose tiling \cite{Collins2017}. Real-space scanning probes like the quantum twisting microscope \cite{Ilani2023} offer another approach by gating the vdW device on a rotating platform, thereby enabling a rotated moir\'{e} effect. The optical conductivity, absorption coefficient and the photocurrent which depend on the interband transitions \cite{Yingying2010,Yin2016,DArora2023}, and experiments analogous to  cryogenic nanoscale photovoltage measurements \cite{hesp2023cryogenic} (Appendix-\ref{ssec:10}) on hBN-encapsulated TBLG may also exhibit the MF signatures.
Future research will investigate MF symmetry \cite{Yazdani2021}, implications for strong-correlation physics, extension to external potentials without common moir\'{e} periodicity \cite{Koshino2021,Koshino2022,Mjat2023}, any nontrivial topological properties embedded in our incommensuration measure, and the possibility of fractality in moir\'{e} quasicrystals \cite{Yao2018q,Moon2019Quasi,Crosse2021,Hammer2022,Aviuri2023superconductivity}.

\begin{acknowledgments}
SG is supported by MTR/2021/000513 funded by SERB, DST, Govt. of India. DA is supported by a UGC fellowship.
\end{acknowledgments}

\appendix
\section{The properties of the fractal generators (FG)}
\label{ssec:1}
Here we provide details of the transformation mappings for FGs corresponding to $ q=3 $, $ p=1 $ and for $q=2, p=1$, respectively, whose effect on the band structure was shown in FIG. (2) of the main text. More examples with additional discussion are provided in Table-\ref{table:generators}. The first case corresponds to the commensurate angle $\theta \sim \ang{21.79} $ which has recently been explored experimentally for pristine TBLG \cite{Ilani2023}. The number of contraction mappings, that is, an associated cardinal number $ n_{c} $ of IFS $ W = \left\{ w_{n}:n=1,2,\dots,n_{c}\right\} $ \cite{Duvall1992} is obtained for each $ L_{N} $ and $ \beta $. In this case $L_{N}$ is found to be $7$ and lie at the intersection of $ x = 1 $ and $\beta = 1 $ lines as shown in FIG. 1(a) of the main text. Correspondingly, $ n_{c} $ of the IFS comes out to be $3$, giving $ W = \left\{w_{1}, w_{2} , w_{3}\right\} $, where $w_{1} = \mathcal{R}(-\phi_{1})\frac{\mathcal{I}}{s},~
w_{2} = \mathcal{R}(\phi_{2})\frac{\mathcal{I}}{s} + w_{1},~
w_{3} = w_{1} + w_{2}$
with $s = \sqrt{L_{N}} = \abs{\bm{b}^{(j-1)}_{1/2}}/\abs{\bm{b}^{(j)}_{1/2}} $ being the contractivity factor and $\mathcal{I}$ is a $2 \times 2$ identity matrix. For $ \{p_{1}, p_{2}, s \}= \{1, 2,  \sqrt{7} \}$, $ \phi_{1} = \cos[-1](\frac{2p_{1}+p_{2}}{2s}) $ and $ \phi_{2} = \frac{\pi}{3} - \phi_{1}$. $ D_{f} $ of the attractor $ A $ is obtained using $ L_{N} $ as $ D_{f} = \log(n_{c})/\log(s)$ \cite{Mandelbrot1983}.

\begin{figure}[b]
	\centering
	\includegraphics[scale=0.26]{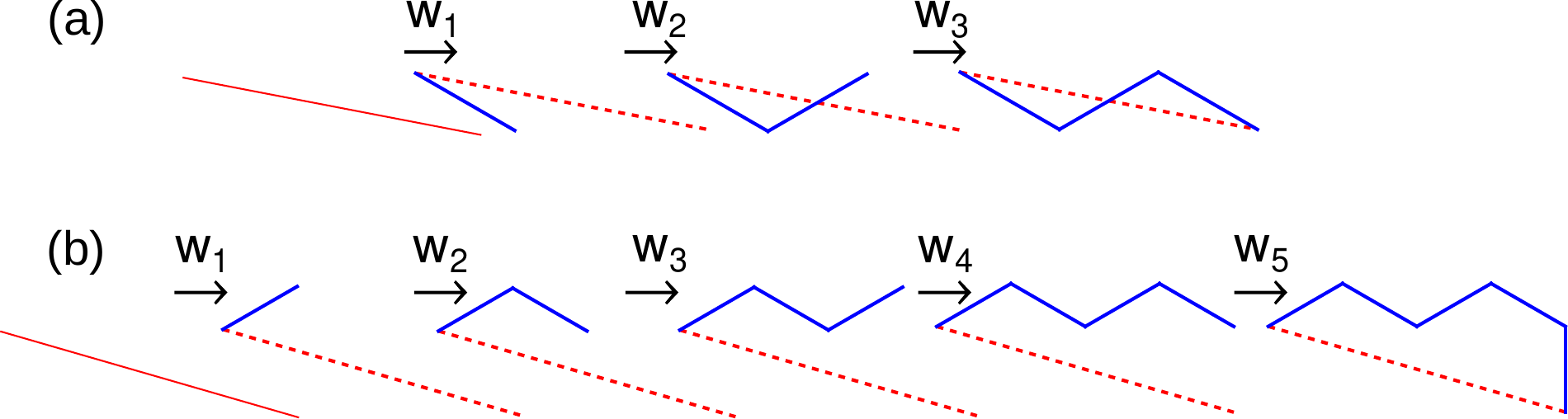}
	\caption{(a) The stepwise creation of the FG for $ L_{N} = 7 $ with $ q=3 $ and $ p=1 $, and (b) $ L_{N} = 13 $ with $ q=2 $ and $ p=1 $. The red dashed line is the side $\bm{u}$ of $A_{0}$ as defined in the Sec.~\ref{ssec:1}}
	\label{fig:FG3121}
\end{figure}

To understand $w_{n} $'s action on one of the sides of $A_{0}$ (BZ of SLG), rotated clockwise with the angle $ \theta/2 \sim \ang{10.89}$, and represented by reciprocal lattice vector $\bm{u} = \frac{1}{3}\left( 2\bm{b}^{-\theta/2}_{1} + \bm{b}^{-\theta/2}_{2} \right) $ (see FIG.~\ref{fig:FG3121}), where $ \bm{b}^{-\theta/2}_{1} $ and $ \bm{b}^{-\theta/2}_{2} $ are the two reciprocal lattice vectors of the rotated graphene layers, we note that $ w_{1} $ shortens $ |\bm{u}|$ by $ s = \sqrt{7} $ times and rotates $\bm{u}$ clockwise by an angle $ \cos[-1](2/\sqrt{7}) $, providing the first side of the FG, namely $\bm{u}_{1} $. Similarly, the successive mappings are $\bm{u}_{2,3} = w_{2,3}\bm{u}$.
The second case mentioned in the main text, corresponds to the commensuration $ \theta \sim \ang{32.20} $ with $ q=2 $ and $ p=1 $, for which $ L_{N} = 13 $ and $ n_{c}=5 $. Considering the side of the initiator to be identical to $ \bm{u}$, the mappings become
$w_{1} = \mathcal{R}\left(\phi_{1}\right)\frac{\mathcal{I}}{s};~~w_{2} = \mathcal{R}(-\phi_{2})\frac{\mathcal{I}}{s} + w_{1} ;~~
w_{3} = w_{1}+w_{2};~~ w_{4}=w_{2} + w_{3}; ~~w_{5} = \mathcal{R}(-(\frac{\pi}{3} +\phi_{2}))\frac{\mathcal{I}}{s}  + w_{4} $
where $ p_{1} = 1 $, $ p_{2} = 3 $ and $ s = \sqrt{13} $. The application of the contraction mappings $ w_{n}\text{'s} $ for the first two entries in Table-(\ref{table:generators}) are shown in FIG.~\ref{fig:FG3121}.

\begin{table}[t]
	\centering
	\caption{Each commensuration is characterized uniquely by a pair of coprime-integers $ \left(q,p\right) $. The L\"{o}schian number $ L_{N} $ and the FG are obtained by identifying the number of sides $ n_{c} $ in the FG. The fractal dimension ($ D_{f} $) corresponding to the various $ q,p $ are calculated using $ L_{N} $ and $ n_{c} $ \cite{Arlinghaus1989}. The outer boundary of the mini-zones superposed over the FBZ of SLG is generated by attaching the FG on a hexagonal initiator.}
	\begin{tabular}{cccccccc}
		\hline
		\hline
		$ \theta $ & $ q $ & $ p $ & $ L_{N} $ & $ \beta $ & $ n_{c} $ & Generator & $ D_{f} $ \\
		\hline
		\hline
		$ \ang{21.79} $ & $ 3 $ & $ 1 $ & $ 7 $ & $ 1 $ & $ 3 $ & 
		\includegraphics[scale=0.12]{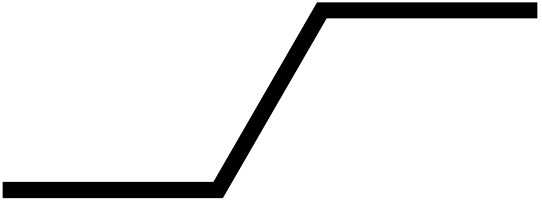} & $ 1.129 $ \\
		$ \ang{32.20} $ & $ 2 $ & $ 1 $ & $ 13 $ & $ 2 $ & $ 5 $ &
		\includegraphics[scale=0.4]{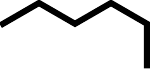} & $ 1.255 $ \\
		$ \ang{13.17} $ & $ 5 $ & $ 1 $ & $ 19 $ & $ 1 $ & $ 5 $ &
		\includegraphics[scale=0.12]{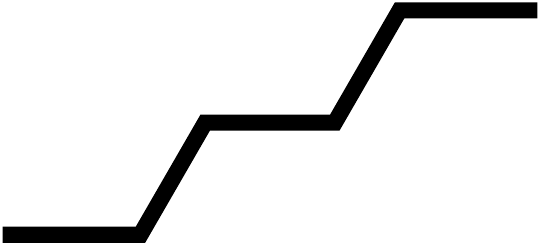} & $ 1.093 $ \\
		$ \ang{38.21} $ & $ 5 $ & $ 3 $ & $ 7 $ & $ 1 $ & $ 3 $ &
		\includegraphics[scale=0.12]{fig18.pdf} & $ 1.129 $ \\
		$ \ang{17.90} $ & $ 11 $ & $ 3 $ & $ 31 $ & $ 4 $ & $ 9 $ &
		\includegraphics[scale=0.42]{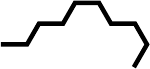} & $ 1.280 $ \\
		\vdots \\
		\hline
		\hline
	\end{tabular}
	\label{table:generators}
\end{table}

\begin{figure}
	\centering
	\includegraphics[scale=0.32]{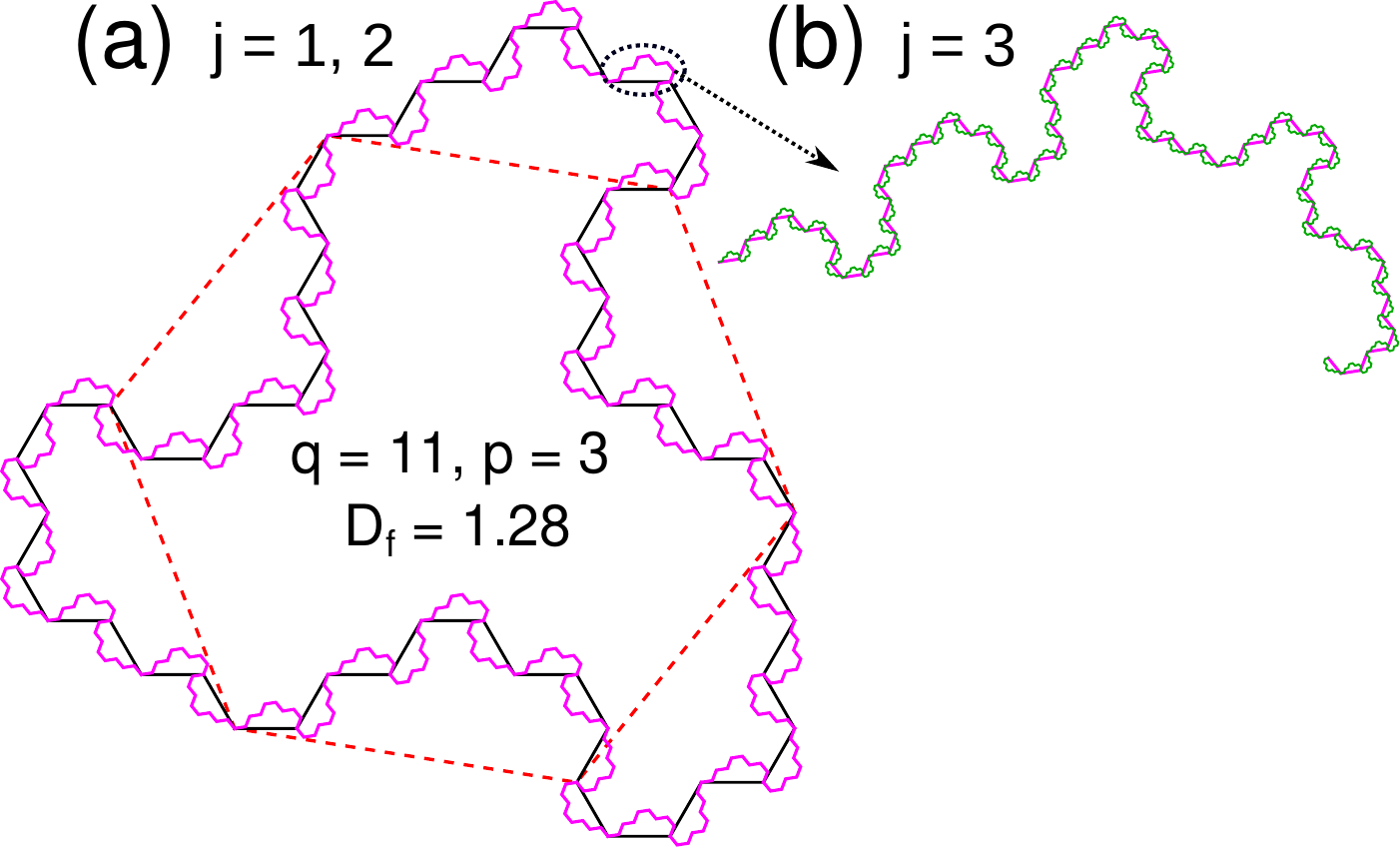}
	\caption{\label{fig:FG113} (a) The fractal corresponding to one more entry in the Table-\ref{table:generators} where $ q=11 $ and $ p=3 $ having $L_{N} = 31 $ with $ \beta = 4 $ at an angle of $ \theta_{r} = \ang{17.89} $ for $ j=1,2 $. (b) the fractal corresponding to the iteration $ j=3 $.}
\end{figure}

The other FGs in Table-(\ref{table:generators}), such as for q=5 and p=3 lead to the same value $L_{N}=7$ and $\beta=1$ as for $q=3$ and $p=1$, but the shift between the Dirac points for commensurate TBLG is $\Delta\bm{K} = \bm{b}^{c}_{2} $. However, the FG remains identical to the one for  $q=3$ and $p=1$. In fact, the mappings for the FGs corresponding to the other values of $ q $ being an odd number with $ p = 1 $ remain the same with an increasing cardinal number $ n_{c} $. Therefore, the class of commensurate structures with $ q $ being an odd number and $ p=1 $ associate with the FGs having the same shape but with a different number of sides. If one draws a line that is perpendicular to the red dashed line in FIG.~\ref{fig:FG3121}(a) which bisects the FG, it exactly cuts it into two pieces with one becomeing the other with a rotation of $ \pi $ in the plane containing the FG. In the case of $q=11$ and $p=3$, the shift $\Delta \bm{K} = \bm{b}^{c}_{2} $ is the same as for $q=5$ and $p=3$ but the FG is asymmetric about the perpendicular bisector. The corresponding FG is shown in FIG.~(\ref{fig:FG113}). It shows the $ j=1,2 $-level iterations in (a) and the one for $ j=3 $ in (b). Similarly, for $q=4$ and $p=1$, the shift in the Dirac points $\Delta \bm{K} = 2\left(2\bm{b}^{c}_{1} + \bm{b}^{c}_{1}\right)/3 $ is very different from the previous cases but the FG has a similar shape as for $q=2$ and $p=1$ where $ \Delta\bm{K} $ is identical.

\section{The real-space construction of the moir\'{e} fractals}
\label{ssec:2}
\begin{figure}
	\centering
	\includegraphics[scale=0.8]{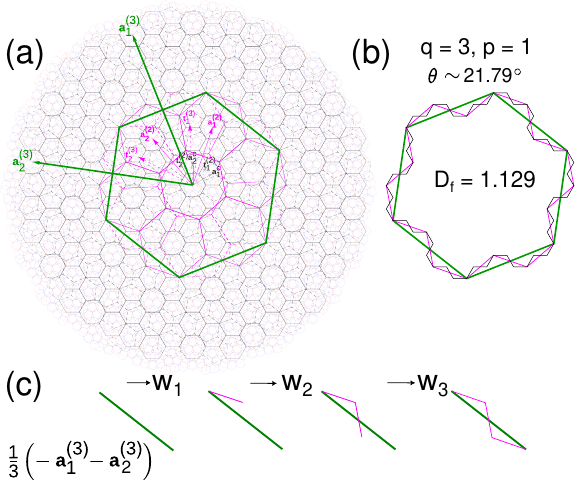}
	\caption{\label{fig:realfracqp31} (a) The real-space structure of commensurate TBLG at $ \theta \sim \ang{21.79} $ and two levels of periodic potentials for $ j=2,3 $. The direct-lattice primitive vectors are also shown for each level. (b) The corresponding moir\'{e} fractal along the edges of the Wigner-Seitz unit-cell at $ j=3 $. (c) The initiator, the arm of the hexagon corresponding to $ j=3 $, and the FG.}
\end{figure}

While the contraction mappings $ \left\{w_{1}, w_{2}, w_{3}, \dots \right\} $ for a moir\'{e} fractal in real-space corresponding to a given value of $ q,p $ remain identical, the initiator however is different. In reciprocal-space, the FG is applied to the side of a hexagon of a rotated SLG in the commensurate case while it's applied to an arm of the moir\'{e} BZ for the incommensurate case to obtain the iterated fractal. However, the initiator in the real-space is the arm of a hexagon which corresponds to the highest-level of iteration of the chosen $ j $-values. \textit{E.g.,} FIG.~\ref{fig:realfracqp31}(b) shows the moir\'{e} fractal corresponding to $ q=3 $ and $ p=1 $. Since the $ j $-values considered are $ 0,1,2,3 $, the initiator is the side of the hexagon corresponding to the commensurate cell of $ j=3 $.

\section{The relation of the number of bands with the fractal dimension ($D_{f}$)}
\label{ssec:3}
As given in the main text, the number of bands for $\theta=\theta_{r}$ within the bandgap of the two lowest bands at the $\Gamma$-point is $4e^{2\ln(n_{c})/D_{f}}-2$ (see FIG.~2(a3) \& (b3) of the main text), while for $\theta\ne\theta_{r}$, the number of bands is $2e^{2\ln(n_{c})/D_{f}}-2$ (see FIG.~4 in the main text). The dependence of the number of bands on the cardinal number $n_{c}$ and the fractal dimension $D_{f}$ is obtained from $ L_{N} $ which is related to $n_{c}$ and $D_{f}$ as
\begin{equation}
	D_{f} = \frac{\ln(n_{c})}{ln(\sqrt{L_{N}})}
\end{equation}
After some rearrangement, the crucial quantity $ L_{N} $ can be written as
\begin{equation}
	L_{N} = e^{2\ln(n_{c})/D_{f}}
	\label{eqn:loschian}
\end{equation}
The number of bands depends upon $L_{N}$ and therefore, it depends upon $n_{c}$ and $D_{f}$ through (\ref{eqn:loschian}).

\section{Derivation of Eq.~(3) of the main text}
\label{ssec:4}
For a general 2D-Bravais lattice case, the direct lattice primitive vectors of the constituting layers $ \bm{a}_1 $ and $ \bm{a}_{2} $ are not necessarily orthogonal, \textit{i.e.,} $ \bm{a}_1\vdot\bm{a}_2 = a_1\,a_2\,\cos(\phi)\ne 0 $ and also $ \abs{\bm{a}_1} \ne \abs{\bm{a}_2} $. Therefore, the general square matrix that maps an integer pair $ \bm{n} = \left\{n_1,n_2\right\} $ to $ \bm{m} = \left\{m_1,m_2\right\} $ is given by
\begin{equation}
	\bm{m} = 
	\left[\cos\,\theta_{r} - \epsilon\frac{\left(\sigma_{z}\cos(\phi)a_1+i\sigma_{y}a_2\right)}{a_1\sin(\phi)}\sin\,\theta_{r}\right]
	\bm{n}
	\label{eqn:commensurate}
\end{equation}
Here $ \epsilon = \text{sgn}\left[\left(\bm{a}_1\times\bm{a}_2\right)_{z}\right] $, $ a_{1/2} = \abs{\bm{a}_{1/2}} $, and $ \phi $ represents the angle between $ \bm{a}_1 $ and $ \bm{a}_2 $.
For the present hexagonal case , $ \abs{a_1} = \abs{a_2} $ and $ \phi = \ang{60} $, the necessary and sufficient condition for the integer solutions $ m_1, m_2, n_1 $ and $ n_2 $ demands the matrix elements to assume only rational values \cite{Shallcross2010}. The commensurate angle $ \theta_{r} $ then becomes
\begin{equation}
	\theta_{r}(q,p) = 2\tan^{-1}(p/\sqrt{3}q)
	\label{eqn:commangle}
\end{equation}
where $ q > p > 0 $. As $ p/q \rightarrow 0 $ gives $ \theta_{r} \rightarrow \ang{0} $ and $ p/q \rightarrow 1 $ gives $ \theta_{r} \rightarrow \ang{60} $. The commensurate structures are distinguished on the basis of $ \delta = \text{gcd}(p,3) $ and the direct lattice primitive vectors are
\begin{equation}
	\begin{bmatrix}
		\bm{a}_{1}^{c} \\ \bm{a}_{2}^{c}
	\end{bmatrix} =
	\begin{bmatrix}
		p_{1} & p_{2} \\
		-p_{2} & p_{1} + p_{2}
	\end{bmatrix}
	\begin{bmatrix}
		\bm{a}_{1} \\ \bm{a}_{2}
	\end{bmatrix}
\end{equation}
where $ p_{1} = \left(3q-p\right)/\gamma $ and $ p_{2} = 2p/\gamma $ for $ \delta = 3 $ and $ p_{1} = \left(q-p\right)/\gamma $ and $ p_{2} = \left(q-p\right)/\gamma $ for $ \delta = 1 $ and the quantity $ \gamma = \text{gcd}\left[3q+p,3q-p\right] $. For both the cases $ \delta =1 $ and $ \delta=3 $, the two elements in first row $ p_1 $ and $ p_2 $ are positive integers $ \mathbb{Z}^{+} $. Corresponding to the direct space primitive vectors $ \bm{a}^{c}_1 $ and $ \bm{a}^{c}_2 $, the reciprocal space primitive vectors $ \bm{b}^{c}_1 $ and $ \bm{b}^{c}_2 $ are defined such that $ \bm{a}^{c}_{i}\vdot\bm{b}^{c}_{j} = 2\pi \delta_{ij} ~\forall~i,j = 1,2$. Then, the number $ L_{N} $ of the BZ hexagons of commensurate cell enclosed within the BZ of SLG formed by $ \left\{\bm{b}_1,\bm{b}_2\right\} $ are,
\begin{equation}
	L_{N} = \frac{\abs{\left(\bm{b}_1\times\bm{b}_2\right)\vdot\hat{\bm{z}}}}{\abs{\bm{b}^{c}_1\times\bm{b}^{c}_2\vdot\hat{\bm{z}}}} =
	p_1^2 + p_2^2 + p_1\,p_2
\end{equation}

\begin{figure}
	\centering
	\includegraphics[scale=0.42]{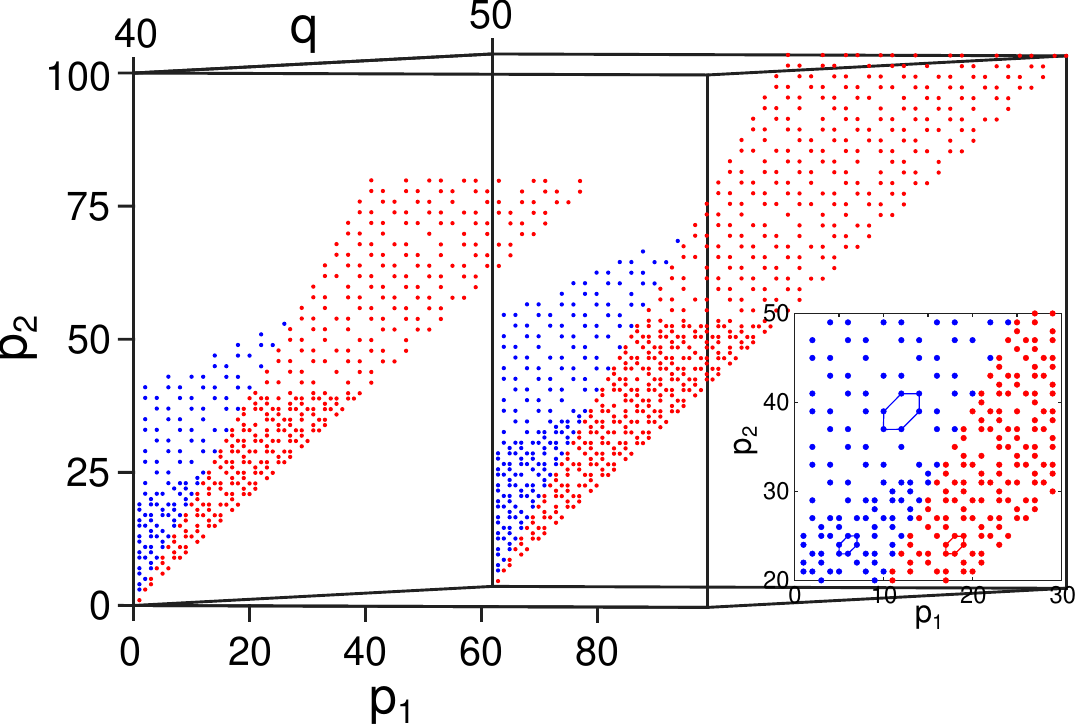}
	\caption{$ p_1 $ vs $ p_2 $ for two maximum values of $q$. Evidently, the points arrange themselves in hexagons and each point is associated with a L\"{o}schian number $ L_N $.}
	\label{fig:p1p2space}
\end{figure}

\subsection{Fractal dimension ($ D_{f} $) in the metric space of $ p_{1}, p_{2} $}
An irregular hexagon is formed for higher magnitudes of $ p_{1}, p_{2} $ along with the smaller hexagons for lower magnitudes. This self-similar, small irregular hexagons lead to the fractality in the metric space $ (\mathbb{Z}^{2},\text{Euclidean}) $. The fractality in the arrangement of the coefficients $ p_{1},p_{2} $ in this space is shown in FIG.~(\ref{fig:p1p2space}). Here we calculate the Hausdorff dimension $ D_{f} $ in the metric space $  (\mathbb{Z}^{2},\text{Euclidean}) $ of coefficients $ p_{1},p_{2} $ with the Euclidean distance. We calculate the $ D_{f} $ using the box-counting theorem \cite{Mandelbrot1983,Lauwerier1991fractals}
\begin{equation}
	D_{f} = \lim_{k\rightarrow \infty}\frac{\log(n_{k})}{\log(2^{k})}
\end{equation}
where $ n_{k} $ is number of smaller polygons that completely fit inside a bigger polygon at the $ k^{\text{th}} $-iteration. There are $ 3^{k} $ small irregular hexagons that fit inside the bigger irregular hexagon at the $ k^{\text{th}} $-iteration and therefore the fractal dimension is then given by
\begin{equation*}
	D_{f} = \frac{\ln(3)}{\ln(2)} \sim 1.585
\end{equation*}
Note that the fractality in this metric space of $ p_{1}, p_{2} $ is different from the fractality that we observed in moir\'{e} fractals where $ D_{f} $ is a function of $ q,p $ that characterizes a given commensuration whereas $ D_{f} \sim 1.585 $ in the metric space of $ p_{1},p_{2} $ is constant.

\section{More on the incommensuration measure}
\label{ssec:5}
It can be shown that for a generic $(q,p)$:
\begin{equation}
	\bm{\Delta K} =
	\begin{cases}
		\frac{2p}{3\gamma}\left(2\bm{b}^{(j)}_{1} + \bm{b}^{(j)}_{2}\right) & \text{if}~\text{gcd}(p,3) = 1 \\
		\frac{2p}{3\gamma}\bm{b}^{(j)}_{2} & \text{if}~\text{gcd}(p,3) = 3 
	\end{cases}
\end{equation}
where $ \gamma = \text{gcd}\left(3q-p,3q+p\right) $ \cite{Mele2010}.
For $q=2n+1$ with $n=1,2,3,\dots$ and $p=1$, the shift $\Delta \bm{K}$ always equals the side-length of the hexagon $ (\ell) $ as shown in FIG.~2(a1) of the main text \cite{Santos2012}. Further, the moir\'{e} lattice vectors coincide with the lattice vectors of the commensurate cell. In this case, we get,
\begin{equation}
	\frac{A^{(j-1)}_{\text{FBZ}}}{A^{(j)}_{\text{FBZ}}} = L_{N} \implies \frac{\Delta A}{A^{(j-1)}_{\text{FBZ}}} = 0
\end{equation}
Similarly, if $q=2n$ with $n=1,2,3,\dots$ and $p=1$, the shift $\abs{\Delta K} = 2\ell$ as shown in FIG.~2(b1) of main text, and the ratio
\begin{equation}
	\frac{A^{(j-1)}_{\text{FBZ}}}{A^{(j)}_{\text{FBZ}}} = \frac{4\chi(q) + 1}{4} \implies \frac{\Delta A}{A^{(j-1)}_{\text{FBZ}}} = \frac{1}{4\chi(q) + 1}
\end{equation}
where $\chi(q)$ is a positive integer dependent on $q$.

\begin{figure}[t]
	\centering
	\includegraphics[scale=0.55]{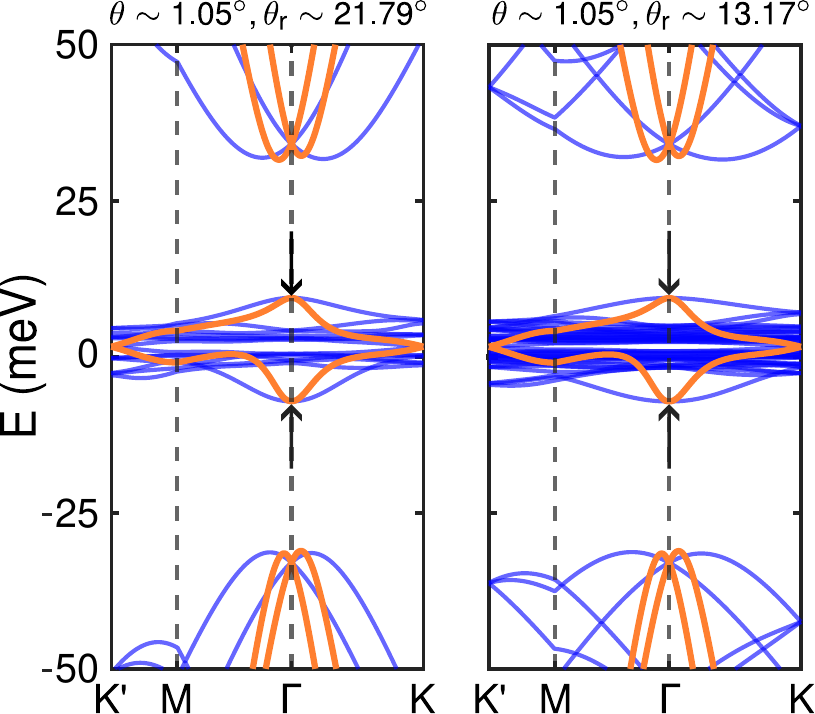}
	\caption{The inclusion of corrugation effects leads to different interlayer hopping parameters, namely, $ t_{\text{AA,BB}} = 79.7~\unit{\milli\electronvolt} $ and $ t_{\text{AB,BA}} = 97.5~\unit{\milli\electronvolt} $ \cite{Koshino2018}. This does not change the $ 2e^{2\ln(n_{c})/D_{f}} - 2 $ bands inserted within the band-gap of the two lowest bands at the $\Gamma$-point (shown by the two arrows) in the absence of mEP and hence the corrugated TBLG also does not affect the emergence of fractality.}
	\label{fig:corrband}
\end{figure}

\begin{figure}
	\centering
	\includegraphics[scale=0.42]{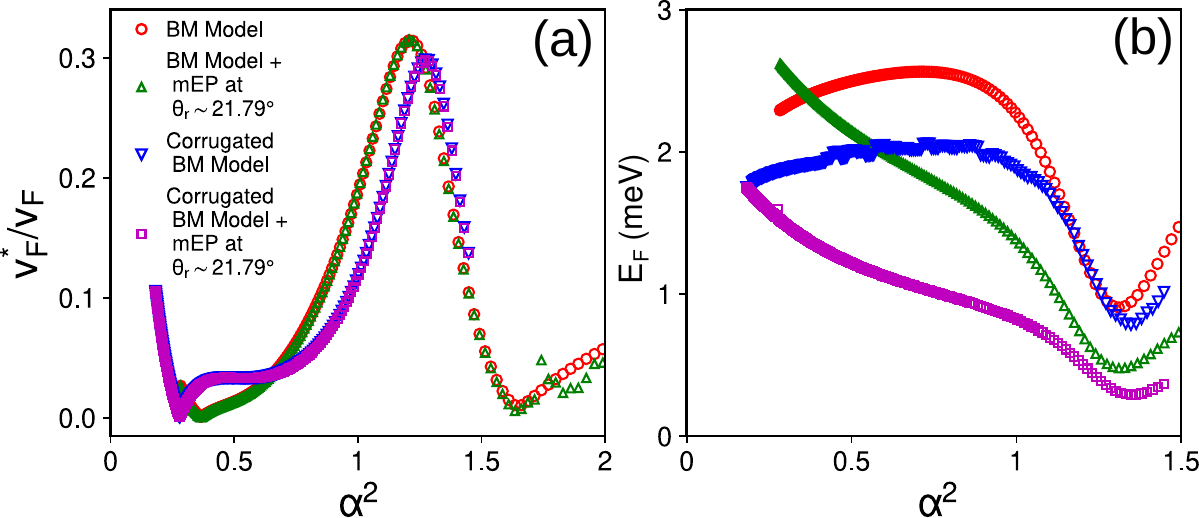}
	\caption{(a) The renormalized Fermi velocity $ v^{*}_{F}/v_{F} $ as a function of $ \alpha^{2} $ and hence the twist angle $ \theta $. (b) The Fermi energy ($E_{F}$) as a function of $ \alpha^{2} $. Similar to the case of a pristine TBLG \cite{Bistritzer2011} or an unrotated mEP to TBLG \cite{Cao2021moireSupflat}, the renormalized Fermi velocity in the presence of a rotated mEP remains unaffected both in the absence and the presence of the corrugation effect.}
	\label{fig:Fermivel1}
\end{figure}

\section{The details of the potential, interlayer tunnelling matrices and the band structures with corrugation effects}
\label{ssec:6}
The cosine potential that we considered in the main text, has a cosine profile, namely
\begin{equation}
	V_{j}(\bmr) = V_0\sum_{i=1}^{3}\cos\left(\bm{G}^{(j)}_{i}\vdot\bmr\right)
\end{equation}
where $ \bm{G}^{(j)}_1 $, $ \bm{G}^{(j)}_2 $ and $ \bm{G}^{(j)}_3 = -\bm{G}^{(j)}_1 - \bm{G}^{(j)}_2 $ are the reciprocal lattice vectors satisfying $ \bm{t}^{(j)}_i\cdot\bm{G}^{(j)}_k = 2\pi\delta_{ik}~\forall~i,k =1,2$. $V_{0}$ is the strength of the potential and the $ \bm{t}^{(j)}_{i} $-vectors are defined in the main text.

The spatially-dependent interlayer-tunneling $ T(\bmr) $ in the Hamiltonian in Eq.~2 of the main text is given by \cite{Bistritzer2011,Vishwanath2019,Ledwith2021}
\begin{widetext}
	\begin{equation}
		T(\bmr) = \sum_{i=1}^{3} T_{i} e^{-i\bm{q}_{i}\vdot\bmr}
		= \sum_{i=1}^{3} \left\{\sigma_{0}t_{\text{AA,BB}} + \left[\sigma_{x}\cos(i-1)\phi + \sigma_{y}\sin(i-1)\phi\right]t_{\text{AB,BA}}\right\} e^{-i\bm{q}_{i}\vdot\bmr}
	\end{equation}
\end{widetext}
where $ t_{\text{AA,BB}} $ and $ t_{\text{AB,BA}} $ are the interlayer hopping parameters in the local AA/BB-regions and AB/BA-regions, respectively. $ \sigma_{0} $ is a second order identity matrix and $ \left(\sigma_{x}, \sigma_{y}\right) $ are the Pauli matrices. The band structures in FIG.~2 and FIG.~4 of the main text are obtained by considering the hopping parameters $ t_{\text{AA,BB}} = t_{\text{AB,BA}} = 110~\unit{\milli\electronvolt} $ \cite{Bistritzer2011} where we ignore the variations in the hopping parameters which occur due to corrugation or atomic relaxations \cite{Carr2018} in different regions. The inclusion of this effect leads to different interlayer hopping parameters in both the regions, namely, $ t_{\text{AA,BB}} = 79.7~\unit{\milli\electronvolt} $ and $ t_{\text{AB,BA}} = 97.5~\unit{\milli\electronvolt} $ \cite{Koshino2018}. The band structures with these values of hopping parameters are shown in FIG.~\ref{fig:corrband}. The emergent fractality remains unaffected and when the fractal dimension is changed from $ D_{f} = 1.129 $ to $ D_{f} = 1.093 $ in going from left to right, an indentical number $ 2e^{2\ln(n_{c})/D_{f}} - 2 $ of bands are inserted within the band-gap of $ \sim 13~\si{\milli\electronvolt} $ at the $\Gamma$-point.

\begin{figure*}
	\centering
	\includegraphics[scale=0.64]{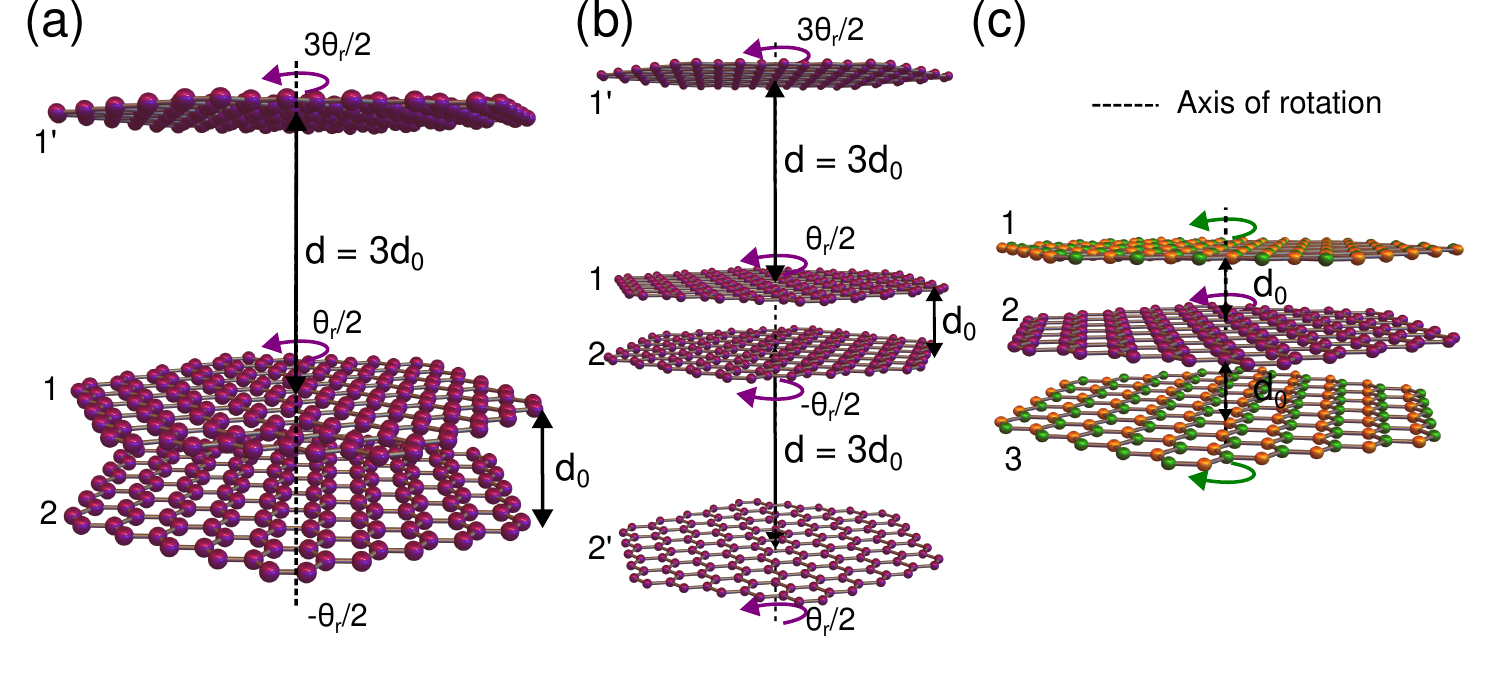}
	\caption{A (a) trilayer-,  (b) a four-layer-twisted graphene, and (c) a trilayer boron nitride-Graphene-boron nitride system. In the above \begin{tikzpicture} \fill[purple] circle (0.15); \end{tikzpicture} denote carbon, \begin{tikzpicture} \fill[orange] circle (0.15); \end{tikzpicture} boron and \begin{tikzpicture} \fill[teal] circle (0.15); \end{tikzpicture} nitrogen atoms, respectively. The angle $ \theta_{r} $ is one of the commensurate angles and $ d_{0} \sim 0.335\,\unit{\nano\meter} $ is the interlayer separation in TBLG. The number besides each layer represents the layer index used in the text in appendix-~\ref{ssec:8}.}
	\label{fig:layersfig}
\end{figure*}
\begin{table*}
	\centering
	\renewcommand{\arraystretch}{.6}
	\begin{tabular}{cc}
		\hline\hline
		System & $H_{\text{pert}}(\bmr)$ \\
		\hline\hline
		Trilayer system &
		$\begin{psmallmatrix}
			\hbar^{2} v^{2}_{F} M^{\text{eff}}_{11'}(\bmr)\sigma_{z} & 0 \\
			0 & \hbar^{2} v^{2}_{F} M^{\text{eff}}_{21'}(\bmr)\sigma_{z} + \mathcal{I}_{2}V^{\text{eff}}_{21'}(\bmr) - \mathcal{I}_{2}V^{\text{eff}}_{11'}(\bmr)
		\end{psmallmatrix}$ \\
		\hline
		Tetralayer system & 
		$\begin{psmallmatrix}
			\hbar^{2} v^{2}_{F} M^{\text{eff}}_{11'}(\bmr)\sigma_{z} + \mathcal{I}_{2}V^{\text{eff}}_{12'}  & 0 \\
			0 & \mathcal{I}_{2}V^{\text{eff}}_{21'}(\bmr) + \hbar^{2} v^{2}_{F} M^{\text{eff}}_{21'}(\bmr)\sigma_{z} + \mathcal{I}_{2}V^{\text{eff}}_{22'}(\bmr) + \hbar^{2} v^{2}_{F} M^{\text{eff}}_{22'}(\bmr)\sigma_{z} - \mathcal{I}_{2}V^{\text{eff}}_{11'}(\bmr)
		\end{psmallmatrix}$\\
		\hline\hline
	\end{tabular}
	\caption{Explicit form of $H_{pert}$ for trilayer and tetralayer graphene systems after ignoring the vector potential term. The quantities $\mathcal{I}_{2}$ is the second order identity matrix and $\sigma_{z}$ is $z$ component of the Pauli matrix. $V^{S}(\bmr)$ in (\ref{eqn:effpot1}) consists of various moir\'{e} potentials $V^{\text{eff}}_{ij}$ between the layers. The integer subscripts $ij$ in  different terms appeared above, refer to the layer indices as indicated in FIG. \ref{fig:layersfig}. The detailed expressions for various terms in $H_{\text{pert}}$ each super-moir\'{e} system.
		are given in the text.}
	\label{table:pert}
\end{table*}

\section{Some realistic systems to realize the model Hamiltonian (2) of the main text at the first iteration of the potential}
\label{ssec:7}
To supplement our assertion in the main text that the moir\'{e} fractals introduced through the model Hamiltonian $H_{2}(\bmr)$ in Eq.(2) of the main text can provide a general description of a number of realistic super-moir\'{e} systems, we shall explicitly provide the modelling for three representative super-moir\'{e} systems in terms of moir\'{e} fractals. For the first iteration $j=2$ of the mEP, the super-moir\'{e} Hamiltonian $H(\bmr)$ can be  written as a combination of the hamiltonian of the moir\'{e}-fractal and a weak periodic perturbation as,
\begin{equation}
	H(\bmr) = H_{2}(\bmr) + H_{\text{pert}}(\bmr)
	\label{eqn:superMF}
\end{equation}
where the periodic perturbation $H_{\text{pert}}(\bmr)$  satisfies 
\begin{equation}
	H_{\text{pert}}(\bmr + n_{1}\bm{a}^{(2)}_{1}+n_{2}\bm{a}^{(2)}_{2}) = H_{\text{pert}}(\bmr)
\end{equation}
where $\bm{a}^{(2)}_{i}$ for $i=1,2$ are the primitive lattice vectors of the super-moir\'{e} cell as defined in the main text. To do this let us note that it has already been established \cite{MKinder2011,MKinder2012} that for moir\'{e} system such as graphene/graphene or graphene-hBN, the full Hamiltonian can be  written as an effective one-layer Hamiltonian experiencing a moir\'{e}-periodic potential as,
\begin{equation}
	V_{\text{moir\'{e}}} = V^{\text{S}}(\bmr)\mathcal{I}_{2} + 
	\hbar^{2}v^{2}_{F} M^{\text{eff}}(\bmr)\sigma_{z} + 
	\hbar v_{F}e\bm{A}^{\text{eff}}(\bmr)\cdot\bm{\sigma}
	\label{eqn:effpot1}
\end{equation}

\begin{figure}[b]
	\centering
	\includegraphics[scale=0.35]{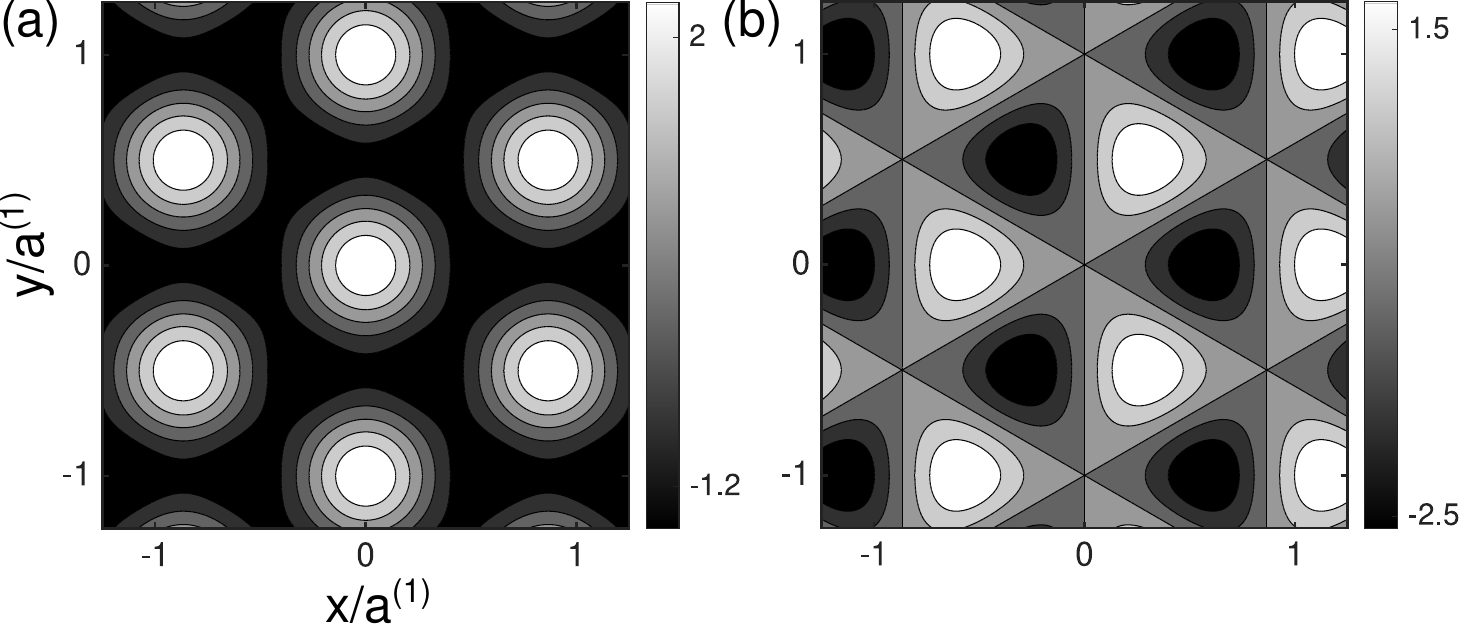}
	\caption{(a) The spatial variation of the inversion symmetry preserving part $ V^{\text{eff}}(\bmr) $, where the colorbar is in the units of $ t^{2}_{0}/V_{\text{STM}} $, and (b) the spatial variation of the mass-dependent term $ M^{\text{eff}}_{11'}(\bmr) $ that breaks the inversion symmetry. Here the colorbar is in the units of $ \sqrt{3}t^{2}_{0}/V_{\text{STM}} $.}
	\label{fig:VeffMeff}
\end{figure}
In the  left hand side (LHS) of Eq. \ref{eqn:effpot1} the strength of each term is of the same order, and is directly proportional to the square of the interlayer hopping parameter and can be controlled by the interlayer bias $V_{\text{STM}}$ (its inversely proportional) which is typically of the order of the bias applied to the tip of the scanning tunnelling microscope (STM) that ranges from $20-500\,\unit{\milli\electronvolt}$ \cite{Choi2019,Nuckolls2023}. $V^{\text{S}}(\bmr)$ preserves the inversion symmetry, the effective mass term $M^{\text{eff}}(\bmr)$ breaks the inversion symmetry while the effective vector potential $\bm{A}^{\text{eff}}(\bmr)$ represents a pseudo magnetic field. In the rest of the calculation, we ignore the effective vector potential (\ref{eqn:effpot1}) since it does not change the proposed insertion of in-gap bands. Thus the first two terms of the LHS of Eq. \ref{eqn:effpot1} form a $H_{\text{pert}}$ whose details for both the tri- and tetra-layer super-moir\'{e} graphene systems are given in Table-\ref{table:pert}. In the following we provide the specifics for each super-moir\'{e} system.

\subsection{A graphene trilayer system}
We consider an AAA-stacked trilayer system \cite{Bernevig2019, Kaxiras2020, Aviuri2023superconductivity} such that the top graphene layer is at a distance $ d = 3\,d_{0} \sim 1\,\unit{\nano\meter} $ where $d_{0}$ is the interlayer perpendicular distance between the remaining two layers. The configuration is shown in FIG.~\ref{fig:layersfig}(a), where the top layer is rotated to an angle $3\theta_{r}/2$, the middle layer is rotated to $ \theta_{r}/2 $ while the bottom layer is rotated at an angle $ -\theta_{r}/2 $ such that the relative misorientation between any two layers is $ \theta_{r} $. Due to the relatively large distance ($ d/d_{0} > 1 $), the top layer couples only weakly with the remaining two layers. The details of the various terms that appeared in the corresponding Hamiltonian $H$ as had appeared in \ref{eqn:superMF} are as follows:
\begin{equation}
	H_{2}(\bmr) = 
	\begin{pmatrix}
		h_{1}(\theta_{r}/2) + V^{\text{eff}}_{11'}(\bmr) & T_{12}(\bmr) \\
		T^{\dagger}_{12}(\bmr) & h_{2}(-\theta_{r}/2) + V^{\text{eff}}_{11'}(\bmr)
	\end{pmatrix}
	\label{eqn:Hamtri1}
\end{equation}
The integer subscripts $ij$ that appeared in  different terms in the Hamiltonian (\ref{eqn:Hamtri1}), again refer to the 
layer indices as indicated in FIG. \ref{fig:layersfig}. These terms, as well as the terms that appeared in $H_{\text{pert}}$ are shown in Table-\ref{table:pert} have the detailed expression as follows:
\bse
\begin{align}
	V^{\text{eff}}_{l1'}(\bmr) & = \frac{6t^{2}_{l1'}}{V_{\text{STM}}} + \frac{t^{2}_{l1'}}{V_{\text{STM}}}
	\sum_{j=1}^{3} \cos(\bm{G}^{l1'}_{j}\cdot\bmr) \\
	\hbar^{2} v_{F}^{2} M^{\text{eff}}_{l1'}(\bmr) & = \frac{\sqrt{3}t^{2}_{l1'}}{V_{\text{STM}}} \sum_{j=1}^{3} \sin(\bm{G}^{l1'}_{j}\cdot\bmr)
\end{align}
\ese
For the moir\'{e} reciprocal lattice vector, the superscript $lj$ indicates the layers involved and it's numbered according to FIG.~\ref{fig:layersfig}, with $l=1,2$.  The subscript numbers such reciprocal lattice vectors between the two surfaces and 
$\bm{G}^{l1'}_{3} = -\bm{G}^{l1'}_{1}-\bm{G}^{l1'}_{2}$. $ t_{11'} = t_{AA/BB}(d) = t_{AB/BA}(d) \sim 7.31\,\unit{\milli\electronvolt}$, $ t_{21'} = t_{AA/BB}(d+d_{0}) = t_{AB/BA}(d+d_{0}) \sim 1.41\,\unit{\milli\electronvolt} $. As a representative value, for $ V_{\text{STM}} = 40\,\unit{\milli\electronvolt} $, the strength of the potential becomes $ t^{2}_{11'}/V_{\text{STM}} \sim 1.334\,\unit{\milli\electronvolt} $ and $ t^{2}_{21'}/V_{\text{STM}} \sim 0.05\,\unit{\milli\electronvolt} $.

\begin{figure}[b]
	\centering
	\includegraphics[scale=0.6]{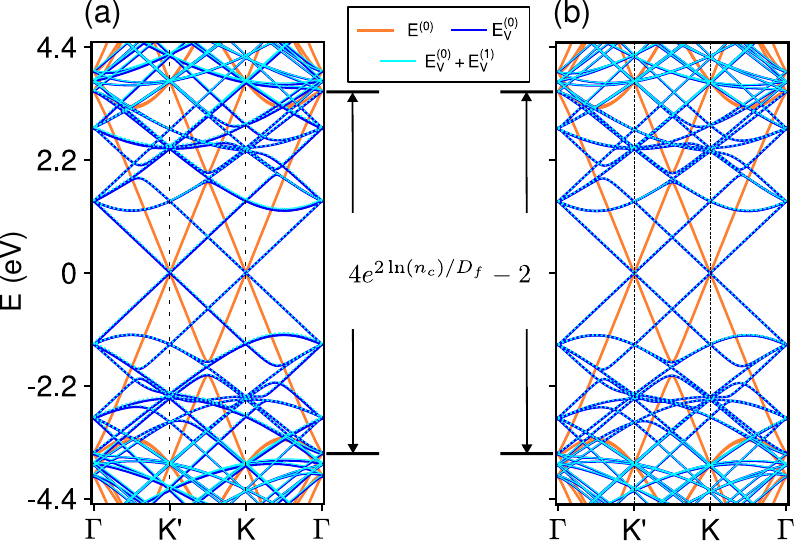}
	\caption{The band structure corresponding to the trilayer graphene system with increasing values of the interlayer bias voltage (a) $V_{\text{STM}} = 10\,\unit{\milli\electronvolt}$, and (b) $V_{\text{STM}} = 40\,\unit{\milli\electronvolt}$. The solid orange lines show the band structure in the absence of the potential while the blue lines represent the eigenvalues of $H_{2}(\bmr) $ without the perturbation, while the cyan lines show the band structure of $H_{2}(\bmr) + H_{\text{pert}}(\bmr) $ calculated upto the first order in perturbation. The dotted cyan lines show $4e^{2\ln(n_{c})/D_{f}}-2$ bands within the bandgap of lowest two bands at the $\Gamma$-point.}
	\label{fig:bstriqp31}
\end{figure}

To proceed further with the calculation we note that for a commensurate angle $ \theta_{r} $ corresponding to $q,p$ in (\ref{eqn:commangle}), the two coprime integers for $2\theta_{r}$ are $q', p'$ that can be obtained from $q,p$ as $ p' = 6pq/\text{gcd}(6qp,3q^2-p^2) $ and $ q' = (3q^2-p^2)/\text{gcd}(6qp,3q^2-p^2) $. After doing straightforward algebra, one can obtain the relation between the reciprocal lattice vectors of two interfaces as
\begin{equation}
	\begin{pmatrix}
		\bm{G}^{11'}_{1} \\ \bm{G}^{11'}_{2}
	\end{pmatrix}=
	\begin{pmatrix}
		Z_{1}(q,p) & Z_{2}(q,p) \\
		Z_{3}(q,p) & Z_{4}(q,p)
	\end{pmatrix}
	\begin{pmatrix}
		\bm{G}^{21'}_{1} \\ \bm{G}^{21'}_{2}
	\end{pmatrix}
	\label{eqn:relation12reci}
\end{equation}
where $Z_{1},\dots,Z_{4}$ are integers and are functions of $q,p$. As an example, for $ q=3 $ and $ p=1 $, the two coprime integers are obtained as $ q'=13 $ and $ p'=9 $ yielding 
$Z_{1}=1, Z_{2}=-1, Z_{3}=1$ and $Z_{4}=2$, in (\ref{eqn:relation12reci}. With these, we calculate the first order correction to the energy eigenvalues of $H_{2}(\bmr)$ sue to $H_{\text{pert}}(\bmr)$. The unperturbed eigenvalues of $H_{2}(\bmr)$ and the corrected eigenvalues to the first order in perturbation $H_{\text{pert}}$ are shown in FIG.~\refeq{fig:bstriqp31} for $ q=3 $ and $ p=1 $ for different values of $V_{\text{STM}}$. From the band structure, we see that the first order corrections to the energy eigenvalues of $H_{2}(\bmr)$ become smaller as the interlayer bias $V_{\text{STM}} $ increases. Therefore, we can conclude that the number of bands $4e^{2\ln(n_{c})/D_{f}}-2$ do not change under the effect of the perturbation.

\subsection{A graphene tetralayer system \cite{Burg2022,Zhang2022}}
The second example of the super-moir\'{e} structure that we consider is where TBLG is sandwiched between two graphene layers, each of which lie at a distance $ d = 3d_{0}$ from it as shown in FIG.~\ref{fig:layersfig}(b). For this system too the unperturbed Hamiltonian can be presented as $H_{2}(\bmr)$ in (\ref{eqn:Hamtri1}). For the expression of $H_{\text{pert}}$ we only retain the interlayer coupling between the nearest-neighbor layers which are dominant over the strengths of the interlayer potential in the next-nearest neighbor coupling. 
Therefore, we ignore $ V^{\text{eff}}_{12'}(\bmr) $, $ V^{\text{eff}}_{12'}(\bmr) $ and $ \hbar^{2} v_{F}^{2} M^{\text{eff}}_{21'}(\bmr) $ in the $H_{\text{pert}}$ in Table-\ref{table:pert}, and $H_{\text{pert}}$ finally becomes
\begin{widetext}
	\begin{equation}
		H_{\text{pert}}(\bmr) = 
		\begin{pmatrix}
			\hbar^{2} v^{2}_{F} M^{\text{eff}}_{11'}(\bmr)\sigma_{z}  & 0 \\
			0 & V^{\text{eff}}_{22'}(\bmr) + \hbar^{2} v^{2}_{F} M^{\text{eff}}_{22'}(\bmr)\sigma_{z} - V^{\text{eff}}_{11'}(\bmr)
		\end{pmatrix}
		\label{eqn:Hamquad2pert}
	\end{equation}
\end{widetext}
Explicit form of different terms in $H_{2}(\bmr)$ and $H_{\text{pert}}(\bmr)$ can be written as 
\bse
\begin{align}
	V^{\text{eff}}_{ll'}(\bmr) & = \frac{6t^{2}_{ll'}}{V_{\text{STM}}}
	+ \frac{t^{2}_{ll'}}{V_{\text{STM}}}
	\sum_{j=1}^{3} \cos(\bm{G}^{ll'}_{j}\cdot\bmr)
	\\
	\hbar^{2} v_{F}^{2} M^{\text{eff}}_{ll'}(\bmr) & = \frac{\sqrt{3}t^{2}_{ll'}}{V_{\text{STM}}} \sum_{j=1}^{3} \sin(\bm{G}^{ll'}_{j}\cdot\bmr)
\end{align}
\ese
The different superscripts and subscripts used have the same meaning as in the previous case. With the $H_{\text{pert}}$ given in (\ref{eqn:Hamquad2pert}) we calculate the first order correction to the energy eigenvalues. The unperturbed and the perturbed band structure to the first order in perturbation theory are shown in FIG.~\refeq{fig:bstriqp31} for $ q=3 $ and $ p=1 $ for comparison. Evidently, the number of bands $4e^{2\ln(n_{c})/D_{f}}-2$ do not change under the effect of the perturbation up to the leading order correction.

\begin{figure}
	\centering
	\includegraphics[scale=0.52]{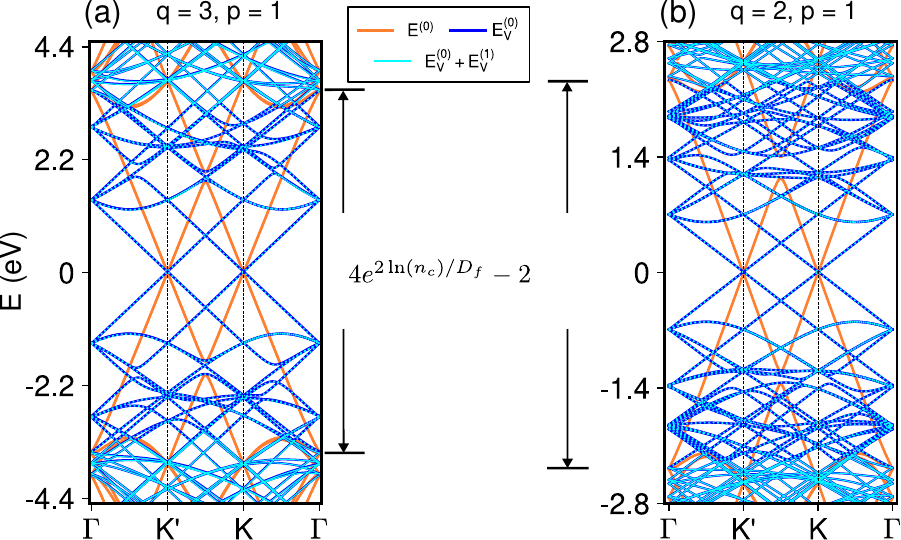}
	\caption{The band structure of the tetralayer graphene system as shown in FIG.~\ref{fig:layersfig}(b) for $V_{\text{STM}} = 20\,\unit{\milli\electronvolt}$. The solid orange lines show the band structure in the absence of the potential while the blue lines represent the eigenvalues of $H_{2}(\bmr) $ without the perturbation{, while} the cyan lines show the band structure of $H_{2}(\bmr) + H_{\text{pert}}(\bmr) $ calculated up to the first order in perturbation. The dotted cyan lines show $4e^{2\ln(n_{c})/D_{f}}-2$ bands within the bandgap of lowest two bands at the $\Gamma$-point.}
	\label{fig:enter-label}
\end{figure}

\subsection{A trilayer system of dissimilar layers}
To show that the modelling of a super-moir\'{e} structure with the Hamiltonian $H_{2}(\bmr)$ of moir\'{e} fractal and a periodic perturbation holds beyond merely multi-layer graphene systems, we consider a system of dissimilar layers \cite{Francois2020} such that a graphene layer is sandwiched between two hexagonal boron nitride (hBN) layers which are a distance $d_0$ apart as shown in FIG.~\ref{fig:layersfig}(c). This system can be modelled as a moir\'{e} fractal at the first iteration of the potential without any perturbing potential, but with a modified $H_{2}(\bmr)$. Namely, 
\begin{equation}
	H_{2}(\bmr) \rightarrow H_{2}'(\bmr) = -i\hbar v_{F}\bm{\sigma}_{\theta_{r}/2}\cdot\bm{\nabla} + V_{21}(\bmr) + V_{23}(\bmr)
	\label{eqn:Hambn2}
\end{equation}
where $V_{2l}(\bmr)$ is the effective periodic potential due to the hBN layer-$ l=1,3 $ on the graphene layer-$2$. This is in contrast to the examples 1 and 2 above since the Hamiltonian $H_{2}'(\bmr)$ itself describes the moir\'{e}-fractal at the first iteration and the perturbation is zero. To progress further, we first find the relative misorientation between the hBN and graphene-layer using
\begin{equation}
	\theta = \sin^{-1}[\left(1+\delta \right)\sin(\phi)] - \phi
\end{equation}
such that they make a commensurate angles $\theta_{r}$ between them \textit{i.e.,} we want the difference $\abs{\phi_{1}-\phi_{2}}$ to be one of the commensurate angle $\theta_{r}(q,p)$ where $\phi$ is the orientation of the moir\'{e}-pattern with the graphene layer. Each potential consists of three terms \cite{MKinder2012}, where the different spatially-dependent terms are given as
\begin{widetext}
\bse
\begin{align}
	V^{\text{eff}}_{2l}(\bmr) & = -3t^{2}_{0} \left(\frac{1}{V_{\text{N}}} + \frac{1}{V_{\text{B}}}\right)
	- t^{2}_{0}e^{-i\psi} \left(\frac{1}{V_{\text{N}}} + \omega \frac{1}{V_{\text{B}}}\right) \sum_{l=1}^{3} \cos\alpha^{2l}_{l}(\bmr)
	\\
	\hbar^{2} v^{2}_{F} m^{\text{eff}}_{2l}(\bmr) & = -\sqrt{3} t^{2}_{0}e^{-i\psi} \left(\frac{1}{V_{\text{N}}} + \omega\frac{1}{V_{\text{B}}}\right) \sum_{l=1}^{3} \sin \alpha^{2l}_{l}(\bmr)
	\\
	\hbar e v_{F} \bm{A}^{\text{eff}}_{2l}(\bmr) & = -2t^{2}_{0}e^{-i\psi} \left(\frac{1}{V_{\text{N}}} + \omega\frac{1}{V_{\text{B}}}\right) \sum_{l=1}^{3} \left\{\cos[\phi(l+1)] \hat{\bm{x}} + \sin[\phi(l+1)] \hat{\bm{y}}\right\} \cos\alpha^{2l}_{l}(\bmr)
\end{align}
\ese
\end{widetext}
where $t_{0} = 0.152\,\unit{\electronvolt}$, $\omega = e^{i\,2\pi/3}$, $\psi \sim -0.29\,\unit{\radian}$, $V_{\text{N}} = 3.34\,\unit{\electronvolt}$ and $V_{\text{B}} = -1.4\,\unit{\electronvolt}$ for boron and nitride atoms \cite{Pmoon2014}, and 
\begin{equation}
	\alpha^{2l}_{l} = \bm{G}^{2l}_{l}\cdot\bmr + \psi + 2\pi/3
\end{equation}
\begin{figure}
	\centering
	\includegraphics[scale=0.48]{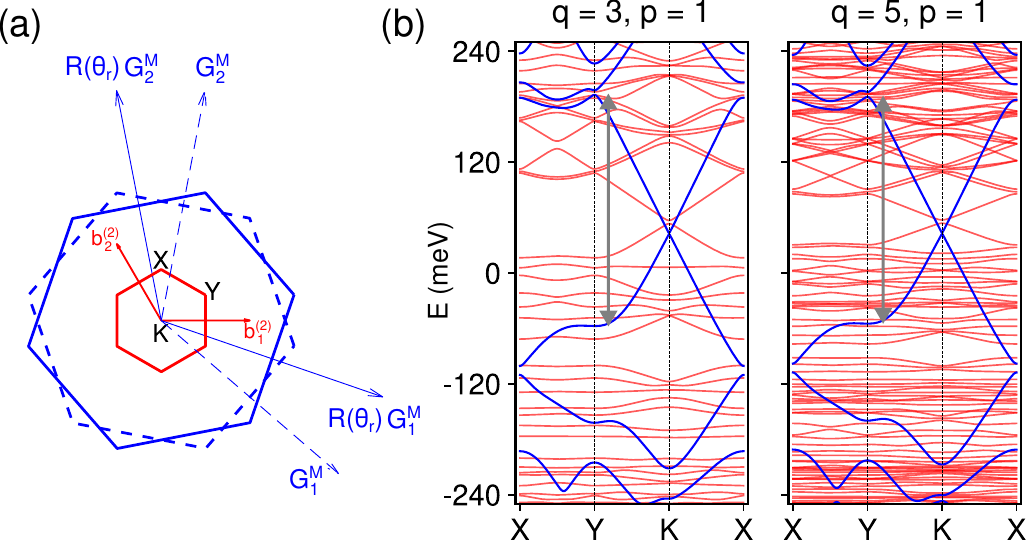}
	\caption{(a) The two blue and dashed blue hexagons show the moir\'{e} BZ of the top and bottom moir\'{e} interfaces for the trilayer hBN-G-hBN shown in FIG.~\ref{fig:layersfig}(c). The inner red hexagon is the super-moir\'{e} BZ. (b) The band structures for two different values of $q,p$ along the high-symmetry path $X$-$Y$-$K$-$X$ \cite{Pmoon2014} as shown in (b). The solid blue lines show the band structures of the graphene-hBN system with $V_{2}(\bmr) = 0$ while the solid red lines represent graphene sandwiched between two hBN layers as in FIG.\ref{fig:layersfig}(c). The two dark-gray arrows show the bandgap between two lowest bands at the $Y$-point where the $2e^{2\ln(n_{c})/D_{f}}-1$ bands are inserted.}
	\label{fig:trilayhbnghbn}
\end{figure}

For three different values of $q,p$ and hence the $\theta_{r}$, the band structures along the high symmetry path X-Y-K-X are shown in Fig~\ref{fig:trilayhbnghbn}. It may be noted that in the band structure plot we have chosen the path through the high-symmetry points differently as compared to the one used in the preceding two examples of super-moir\'{e} structures consisting only out of graphene layers. This is in accordance with the convention used in \cite{Pmoon2014}.
The high-symmetry point $Y$ encloses $2e^{2\ln{n_{c}}/D_{f}}-1$ in-gap bands within the bandgap of the lowest two bands shown by the double-headed arrow. This example of dissimilar layers also exhibits the robustness of the insertion of a controlled number of bands determined by the fractal dimension $D_{f}$ of the moir\'{e} fractal. Therefore, the moir\'{e} fractal can also explain the band structure of such a system.

\section{Discussion on the probability-density plot for of the moir\'{e}-fractal wavefunctions given in FIG. 2 and FIG. 4 of the main text}
\label{ssec:8}
The probability density corresponding to the wave functions of the MF at the first iteration $j=2$ for the Hamiltonian (2) of the main text were 
plotted in FIG. 2 (a4) and (b4) and also in the third column of FIG. 4 in the main text. To that purpose we have calculated the spatial variation of the probability density $\rho_{n\bmk}(\bmr) = \abs{\psi_{n\bmk}(\bmr)}^{2}$ of the Bloch states, where $n$ is the band index and $\bmk$ is the Bloch wave vector.
Specifically, we calculated $\rho_{n\bmk}(\bmr)$ corresponding to the conduction band at the Dirac point over the area in real-space covering the first super-moir\'{e}-cell. $ \theta=\theta_{r}(3,1) \sim \ang{21.79} $ and $\theta_{r}(2,1) \sim \ang{32.20}$, the $\rho_{K}(\bmr)$ is shown in the FIGs.~2(a4) \& (b4) of the main text and $\theta \sim \ang{1,05}$ and $\theta_{r}(3,1) \sim \ang{21.79}$ is shown in FIGs.~4(a)and for $\theta \sim \ang{1,05}$ and $\theta_{r}(5,1) \sim \ang{13.17}$ in FIG.~4(b).

For both the plots the WS unit cell is shown with solid black lines (see FIGs.~2 \& 4) that encloses $ 2e^{2\ln(n_{c})/D_{f}} $ local maxima or minima of $ \rho_{c(v)K}(\bmr) $. We numerically verified that this happens for both the high-symmetry $K$ and $M$-points, while the number of maxima or minima enclosed by the WS cell at the $\Gamma$-point is different.

\begin{figure*}
	\centering
	\includegraphics[scale=0.82]{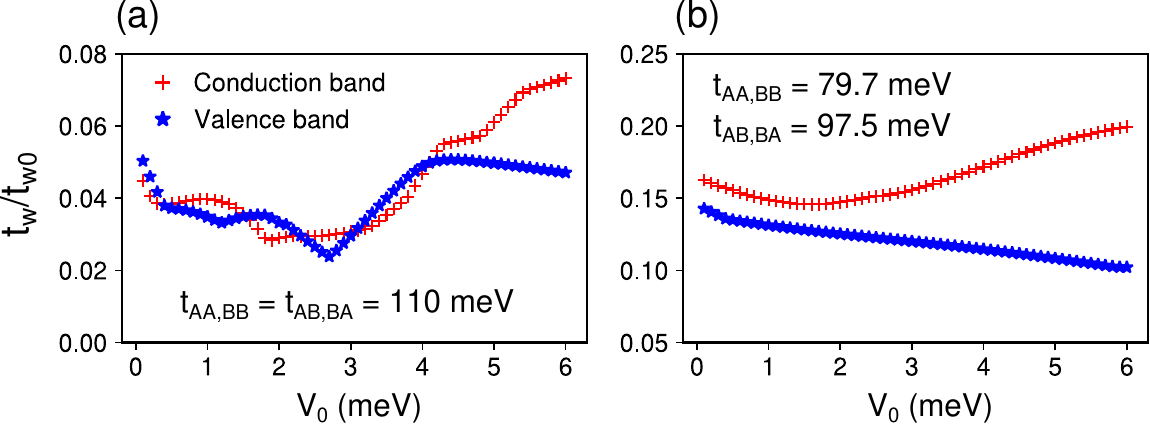}
	\caption{The bandwidth of the conduction and valence band in the presence of mEP for $q=3,p=1$ for (a) $t_{AA,BB} = t_{AB,BA} = 110\,\unit{\milli\electronvolt}$ \cite{Bistritzer2011}, and (b) $t_{AA,BB} = 79.7\,\unit{\milli\electronvolt}$ and $t_{AB,BA} = 97.5\,\unit{\milli\electronvolt}$ \cite{Carr2018,Koshino2018W}. The bandwidth $t_{w}$ is in the unit of $t_{w0}$ which is the bandwidth of pristine TBLG at $\theta \sim \ang{1.05}$.}
	\label{fig:bw}
\end{figure*}

\section{Calculation of the Hubbard parameters}
\label{ssec:9}
The Hubbard interaction $U_{\text{H}}$ can be written as \cite{Coleman_2015}
\begin{multline}
	U_{\bm{R}'m',\bm{R}m} = \\
	\sum_{XX'}\int \dd{\bmr'}\dd{\bmr} \abs{\phi^{X'}_{m'}(\bmr',\bm{R}')}^{2}~U_{\text{C}}(\bmr',\bmr)~\abs{\phi^{X}_{m}(\bmr,\bm{R})}^{2}
\end{multline}
where $U_{\text{C}}(\bmr',\bmr)$ is the screened Coulomb interaction, $e$ is the electronic charge and $\phi^{X}_{{m}}(\bmr,\bm{R})$ is the Wannier orbital of the sublattice $X$ and band index $m$ that is centered at the $\bm{R}$-th lattice site. For pristine TBLG, the localized Wannier orbitals can be constructed from the Bloch states of the Hamiltonian $H_{1}$ in Eq.~(2) of the main text corresponding to the two flat bands near the Fermi level and these orbitals are centered at the local AB/BA-regions of the moir\'{e} pattern \cite{HoiChun2018,Yuan2018,Koshino2018W}.
Following this prescription \cite{Yuan2018}, for the onsite Hubbard interaction $U_{0}$ one can write 
\begin{equation}
	U_{0} \propto \frac{e^{2}}{a^{(1)}}
\end{equation}
where $e$ is the electronic charge and the moir\'{e} wavelength $a^{(1)}$ provides the cut-off for the screening. 
In the moir\'{e} fractal model due to the presence of the mEP in $H_{j}$ for $j \geq 2$ in Eq.(2) of the main text, the Bloch states $\psi^{(j)}_{n\bmk}(\bmr)$ corresponding to the $j$th iteration of the potential have the Bloch periodicity corresponding to the 
moir\'{e} super-cell, that can be expressed as 
\begin{equation}
	\psi^{(j)}_{n\bmk}(\bmr + \bm{a}^{(j)}) = e^{i\bmk\cdot\bm{a}^{(j)}}\psi^{(j)}_{n\bmk}(\bmr)
	\label{eqn:Blochper}
\end{equation}
The translational-invariant Wannier functions, made out of superposing these Bloch states, will be centered at the lattice sites given by $\bm{a}^{(j)}$. 
Accordingly, the onsite Hubbard interaction $U_{0}$ that has the lattice constant as a cut-off length, will be scaled. 
However it may be mentioned that the Wannier orbitals are constructed through self-consistent \emph{ab-initio} calculations and depend upon the number of chosen bands \cite{Marzari2012,Yuan2018,HoiChun2018}. In a moir\'{e} fractal, $2e^{2\ln(n_{c}/D_{f})}-2$ inner bands can be considered since they are well-separated from the other higher bands (see FIG.~4 in main text and FIG.~5 here). A more detailed calculation may lead to a more precise quantitative estimate of $U_{0}$ in the MF, but this is beyond the scope of the current manuscript. Nevertheless, following the above argument we can estimate the onsite Hubbard interaction for the $j$th iteration of the potential as
\begin{equation}
	U^{(j)}_{0} \propto \frac{e^{2}}{\epsilon\,a^{(j)}}
\end{equation}
If $U_{0}$ is the onsite Hubbard interaction for pristine TBLG, then for $j=2$, it becomes $ U_{0} \rightarrow U_{0}/s $ where $s$ is the contractivity factor as defined in the main text. Particularly, for $q=3$ and $p=1$, the contractivity factor $s=\sqrt{7}$. For the bandwidth $t_{W}$, however we do not have any such simple scaling argument. The bandwidth $t_{W} = \text{max}(E_{n}) - \text{min}(E_{n})$, where $E_{n}$ is the energy of the $n$th band, can be determined as a function of the strength of the potential $V_{0}$. Hence we determine this numerically and present the results in FIG.~\ref{fig:bw}(a) and (b). 
It can be seen that that the band width depends on both the $V_{0}$ as well as the hopping parameters. Even though the band-width gets significantly reduced as compared to pristine TBLG, the full behavior is not amenable to simple explanation. 
As an example, $t_{w}/t_{w0} \sim 0.04$ for the conduction band at $V_{0} = 1.2\,\unit{\milli\electronvolt}$ as shown in FIG.~\ref{fig:bw}(a) for $t_{AA/BB} = t_{AB/BA} = 110\,\unit{\milli\electronvolt}$ \cite{Bistritzer2011}.
Therefore, the Hubbard ratio $U/t_{w}$ becomes $1/(\sqrt{7}\times 0.04) \sim 9.4 $ of the ratio of pristine TBLG.

\section{More on the experimental signatures of moir\'{e} fractals}
\label{ssec:10}
In the presence of the first iteration of an external potential $V_{2}(\mathbf{r})$, the in-gap bands near the Fermi surface, situated within the energy window of the lowest two bands, contribute to a greater number of dips and peaks in the density of states (refer to FIG.~\ref{fig:fractal_structures_comm} and \ref{fig:frac_strs_incomm} of the main text). This effect arises from the curvature of these additional bands, resulting in changes to the density of states within that energy range. Since the differential conductance $ (\dd{I}/\dd{V}) $ in the various real-space probes is proportional to the DOS of the sample at a particular bias volatge $ V_{\text{bias}} $ \cite{Binnig1982},
\begin{equation}
	\dd{I}/\dd{V} \propto \rho_{\text{sample}}(-e\,V_{\text{bias}})
\end{equation} where $e$ is the fundamental electron charge. The increased DOS of the sample leads to an increased conductance within a given energy window. In case of the quantum twisting microscope \cite{Ilani2023}, which offers better real-space probing due to the local interference at the tip, the increased number of states at a particular location $\bmr$ may lead to an enhanced coupling among these states and therefore alter the transport properties in contrast to the case when there is no such external potential.

The calculation of experimentally measurable optical properties involves the calculation of the optical matrix elements between the Bloch states of different (same) band indices, \emph{i.e.}, the interband (intraband) transitions \cite{Yingying2010,Yin2016,DArora2023}. Due to the greater number of states available within a given energy window, there may be non-vanishing optical matrix elements between the induced in-gap states that can further tune those properties and therefore optical measurements also provide a technique for characterizing MFs.

A recent article \cite{hesp2023cryogenic} reported cryogenic near-field optoelectronic measurements of hBN-encapsulated MATBLG where the photovoltage measurements revealed a supermoir\'{e} pattern whose periodicity was embedded in the photovoltage response. In their experiment, the top graphene layer was aligned with the top encapsulating hBN layer, while the bottom graphene layer was twisted by an angle $\theta_{\text{hBN}}$ \emph{wrt} the hBN layer, and the two graphene layers were relatively twisted by $\theta_{\text{TBLG}}$. Since the moir\'e-wavelength for the graphene-hBN interface is limited to $\sim 14\,\unit{nm}$ due to the lattice mismatch of $ 1.8\,\% $, there is a finite mismatch in the moir\'{e}-wavelengths of both the graphene-hBN and the graphene-graphene interfaces at smaller angles. For MFs in hBN-encapsulated graphene (FIG.\ref{fig:trilayhbnghbn}), both the hBN layers with respect to the graphene layer are rotated to the same angle, thereby enabling the same moir\'{e} wavelengths in both the graphene-hBN interfaces. Thus for the two cases with $ (q,p) = (3,1) $ and $ (q,p) = (5,1) $ the supermoir\'{e} wavelengths become $ \lambda_{\text{SM}} \sim 36.13\,\unit{\nano\meter}$ and $ \lambda_{\text{SM}} \sim 60.24\,\unit{\nano\meter}$, respectively. A similar photovoltage response measurement in hBN-encapsulated graphene may also show supermoir\'{e} signatures.

\end{document}